\documentclass{article}

\newtheorem{assumption}{Assumption} 
\usepackage{PRIMEarxiv}
\usepackage{natbib}
\usepackage[utf8]{inputenc} 
\usepackage[T1]{fontenc}    
\usepackage{hyperref}       
\usepackage{url}            
\usepackage{booktabs}       
\usepackage{amsfonts}       
\usepackage{nicefrac}       
\usepackage{microtype}      
\usepackage{fancyhdr}       
\usepackage{graphicx}       
\graphicspath{{media/}}     
\usepackage{amssymb}
\usepackage{booktabs}
\usepackage{siunitx}
\usepackage{float}
\usepackage{amsmath}

\newtheorem{theorem}{Theorem}[section]
\newtheorem{lemma}[theorem]{Lemma}
\newtheorem{proposition}[theorem]{Proposition}
\newtheorem{corollary}[theorem]{Corollary}
\newtheorem{proof}[theorem]{Proof}
\newtheorem{remark}[theorem]{Remark}

\newcommand{\bSigma}{\mbox{\boldmath $\Sigma$}}
\newcommand{\btheta}{\mbox{\boldmath $\theta$}}

\usepackage{threeparttable}

\title{Estimation of Treatment Effects based on Kernel Matching 
\thanks{\textit{\underline{Citation}}: 
\textbf{Authors. Title. Pages.... DOI:000000/11111.}} 
}

\author{
  Chong Ding, Zheng Li, Wei Gao \\
  Key Laboratory for Applied Statistics of MOE, School of Mathematics and Statistics,\\
  Northeast Normal University, \\
  Changchun 130024, China. \\
  \texttt{\{Wei Gao\}gaow@nenu.edu.cn} \\
   \And
  Hon Keung Tony Ng \\
  Department of Mathematical Sciences,  \\
  Bentley University,\\
  Waltham, Massachusetts 02452, U.S.A. \\
 \\
}

\begin{document}
\maketitle

\begin{abstract}
Kernel matching is widely used to estimate treatment effects. This paper establishes the consistency and asymptotic normality of kernel-matching estimators for both the average treatment effect (ATE) and the average treatment effect on the treated (ATT), thereby providing a rigorous theoretical foundation for their use in causal inference. We further show that, under certain structural assumptions on the propensity score, these estimators can attain the semiparametric efficiency bound, consistent with the results of \cite{hahn1998role}.  We also derive the asymptotic distributions of the ATE and ATT estimators when the propensity score is estimated rather than known, extending our theoretical results to the empirically relevant setting. For the ATE, using an estimated propensity score can yield a lower asymptotic variance than using the true propensity score. For the ATT, however, no unambiguous efficiency ordering between the two approaches can be established. Through Monte Carlo simulations, these estimators exhibit competitive finite-sample performance relative to standard treatment-effect estimators. Finally, we illustrate the proposed approach using data from the National Supported Work job-training program.
\end{abstract}



\section{Introduction}
Causal inference has become a cornerstone of modern empirical research, providing a rigorous framework for distinguishing between causal relationships and mere correlations. A fundamental object of interest is the treatment effect, which quantifies the causal impact of a treatment by comparing outcomes between treatment and control groups. Estimating treatment effects is challenging because, for each unit, we only observe one of the two potential outcomes that could have occurred.  The process of estimating causal effects fundamentally depends on imputing a potential outcome for an individual, calculated as a weighted average of the observed outcomes of their nearest neighbors from the relevant counterfactual set \citep{rosenbaum1983central}. Thus, causal effects are typically expressed as the difference between the imputed outcome and the actual observed outcome, similar to what would be expected from a randomized experiment. A variety of matching methods have been developed over time, including Mahalanobis metric matching \citep{rosenbaum1985constructing}, full matching \citep{rosenbaum1991characterization,hansen2004full}, genetic matching \citep{diamond2013genetic}, and methods based on sufficient dimension reduction \citep{luo2020matching}.

As two widely used matching approaches, covariate matching \citep{abadie2006large} and propensity score matching \citep{abadie2016matching} have many strengths, though practical implementation can be demanding because they often call for close matches across all covariates or the propensity score. \cite{imbens2009recent} also noted that guidance on optimal match selection and certain implementation choices remains limited. Consequently, in common implementations, such as nearest-neighbor and propensity score matching \citep{abadie2006large, abadie2016matching}, the number of neighbors must generally be prespecified, and principled data-driven guidance remains limited. Performance can also be affected when matches are relatively distant (e.g., treated units paired with controls with substantially different propensity scores), which may introduce bias if such pairs are retained in estimation \citep{becker2002estimation}. 
Alternative approaches include inverse probability weighting (IPW) estimation and doubly robust (DR) estimation. Note that the IPW estimator can be unstable under extreme propensity scores \citep{hirano2003efficient,imai2014covariate}, and the DR estimation remains consistent if either the propensity score or outcome model is correctly specified \citep{bang2005doubly}.

Kernel matching is a widely used nonparametric approach to treatment-effect estimation that assigns weights to comparison units according to their similarity to treated units \citep{handouyahia2013kernel}. Unlike interval and nearest-neighbor matching, it uses continuous weights that give greater influence to units with more similar covariate values. \cite{smith2005does} provided an intuitive account of this approach, while related extensions include local linear and local quadratic matching \citep{HAM2011208}. By smoothing over the covariate space, kernel matching can reduce sensitivity to discrete match selection and yield stable estimates when there is sufficient overlap between the treatment and control groups. However, applying kernel smoothing directly to a high-dimensional covariate vector may suffer from the curse of dimensionality.
To address this limitation, kernel matching can instead be performed on the propensity score, defined as the probability of receiving treatment conditional on the observed covariates \citep{rosenbaum1983central}. Because the propensity score is a scalar balancing score, this approach reduces the matching problem to one dimension \citep{abadie2016matching}. Kernel weights are then determined by the similarity of units' propensity scores, facilitating comparisons when exact matches are unavailable and improving the comparability of treated and control units.

In this paper, we employ a kernel-matching estimator within a broader methodological framework and investigate its large-sample behavior. We establish consistency and asymptotic normality, contributing to the theoretical understanding of this approach. Although popular in applications, the asymptotic properties of kernel-matching estimators have been less fully developed; our results help clarify these properties.   
\cite{hfecon} showed that bias-corrected nearest-neighbor matching with a diverging number of matches is doubly robust and semiparametrically efficient under suitable conditions. \cite{LIN2025106080} further established analogous properties for a broad class of regression-adjusted imputation estimators.
\cite{hf} studied the asymptotic properties of propensity-score-matching estimators when the number of matches diverges with the sample size. Their results show that matching with a diverging number of matches $M$ can attain the semiparametric efficiency bound under suitable sufficient-dimension-reduction conditions. Analogously, in this work, we show that kernel-matching estimators can also attain this efficiency bound under corresponding sufficient-dimension-reduction assumptions.
In comparison to nearest-neighbor matching, kernel matching provides a smooth, weighting-based alternative that replaces discrete neighbor selection with continuous kernel weights. This can reduce sensitivity to the choice of a fixed number of neighbors and may provide additional flexibility through bandwidth-driven smoothing. At the same time, performance depends on standard tuning decisions (e.g., kernel and bandwidth choices); therefore, we view kernel matching as a useful complement to existing methods rather than a uniformly superior replacement.

This paper makes several main contributions. First, we develop kernel-matching estimators based on an estimated propensity score for both the ATE and the ATT. Whereas \cite{heckman1998matching} considered a local kernel estimator specifically for the ATT and focused primarily on theoretical analysis, we provide a unified and more accessible treatment of kernel matching for both estimands, establish asymptotic normality under weaker assumptions, and evaluate the finite-sample performance of the proposed estimators through extensive simulations and a real-data application. We also systematically examine the practical effects of kernel and bandwidth selection, which were not investigated by \cite{heckman1998matching}. Second, we characterize how propensity score estimation affects the asymptotic efficiency of kernel-matching estimators. For the ATE, the resulting adjustment to the asymptotic variance is non-positive and is zero only under special conditions, showing that matching on an estimated propensity score can be more efficient than matching on the true propensity score. For the ATT, however, the variance adjustment has no definite sign and depends on the underlying data-generating process. This distinction has important implications for inference: ignoring propensity score estimation may yield confidence intervals that are unnecessarily wide or, more seriously, too narrow. Third, we characterize the relationship between kernel matching and the semiparametric efficiency bounds for the ATE and ATT. We show that propensity-score kernel matching attains these bounds when the propensity score provides sufficient dimension reduction for the relevant conditional outcome moments. We further develop doubly robust kernel-matching estimators that attain the corresponding efficiency bounds under a more general sufficient-dimension-reduction structure, and provide consistent influence-function-based variance estimators for statistical inference.

The rest of the paper is organized as follows. In Section \ref{sec:2}, we introduce the mathematical notation and the concepts of treatment effect estimation and describe the kernel matching estimator. In Section \ref{sec:3}, we establish the asymptotic properties of the kernel-matching estimator. Monte Carlo simulation studies are used to evaluate the performance of the estimators and select appropriate bandwidths and kernel functions in Section \ref{sec:4}. We also assess the efficacy of our estimator under model misspecification in Section \ref{sec:4}. 
Section \ref{sec:5} provides a practical data analysis to illustrate the  methodology and a simulation study based on real datasets to assess the performance of estimators under bad and
good overlap conditions.  
Finally, some concluding remarks are provided in Section \ref{sec:6}. 
 
\section{Basic Notation and Concepts of Kernel Matching}
\label{sec:2} 

Following the framework of \cite{rubin1974estimating}, the treatment effect on the potential outcomes can be defined as follows.
For unit $i$ ($i = 1, \ldots, N$), let the variable $W_i$ indicate whether an active treatment was received $(W_{i} = 1)$ or not received $(W_{i} = 0)$, for $W_{i} \in \{0, 1\}$. The observed outcome is $Y_i = W_i Y_i(1) + (1-W_i) Y_i(0)$, where $Y_{i}(1)$ and $Y_{i}(0)$ indicate the potential outcomes that the individual would have received active treatment or not, respectively. In other words, for unit $i$, the observed outcome $Y_i$ for treatment $W_i$ can be expressed as 
$$
Y_i = \begin{cases}Y_i(0) & \text { if } W_i=0, \\ Y_i(1) & \text { if } W_i=1.\end{cases}
$$
 In this paper, we are interested in ATE
\begin{equation*}
\tau = E[Y(1)-Y(0)],     
\end{equation*}
and ATT  
\begin{equation*}
\tau_{t} = E[Y(1)-Y(0) \mid W=1].  
\end{equation*}
The assumptions listed below are essential for establishing limit properties and will remain part of the underlying assumptions throughout this article. These assumptions are commonly found in the existing literature for causal inference  \citep[see, for example,][] {rosenbaum1983central,abadie2016matching} and are considered standard assumptions.
\begin{assumption}
Let $X$ be a $d$-dimensional random vector of covariates and let $\mathbb{X}$  denote the support of $X$. Define the propensity score as $p(x)=\Pr(W=1|X=x)$. We consider 
\begin{itemize} 
\item[{(i)}] (Unconfoundedness) $W\perp \{Y(1), Y(0)\} | X$, where $\perp$ denotes  independence \citep{dawid1979conditional};
\item[{(ii)}] (Overlap) $\underline{\eta}\leq p(X) \leq \overline{\eta}$ almost surely, for some $\underline{\eta}>0$ and $\overline{\eta}<1$.	
\end{itemize}
\label{assum1}
\end{assumption}  
Note that Assumption \ref{assum1} is also known as "strong ignorability" \citep{rosenbaum1983central}, which implies that $X$ is sufficient for eliminating the impact of confounding factors (i.e., "unconfoundedness”). \cite{hahn1998role}  established bounds for the asymptotic variance and explored asymptotically efficient estimation methods under Assumption \ref{assum1}(i).

Next, we define the kernel-matching estimators for the ATE and ATT. 
Following the notation defined above, for $N\in \mathbb{N}$, suppose $\{Y_i, W_i, X_i\} _{i=1}^N$ are independent draws from the population with the distribution of $\{Y,W,X\}$. Let $N_1$ and $N_0$ be the number of treated and control units, respectively. The individual treatment effect is $\tau_i = Y_i(1)-Y_i(0)$. For every unit in the study, only one of the two potential outcomes, $Y_i(0)$ or $Y_i(1)$, is observed, and the other is unobservable or missing.

Let $K(\cdot)$ be a kernel function, and let $h$ be the bandwidth parameter that converges to zero as $N$ goes to infinity. The kernel-matching estimators based on the propensity score $p(X)$ for the missing potential outcomes can be expressed as
$$
\widehat{Y}_i(0)=
\begin{cases}
Y_i & \text{ if } W_i=0, \\
\frac{\sum_{j:W_j=0}^N Y_j K\left(\frac{p(
X_j)-p(X_i)}{h}\right) }{\sum_{j:W_j=0}^N K\left(\frac{p(X_j)-p(X_i)}{h}\right)} & \text { if }  W_i=1,
\end{cases}
$$
and 
$$
\widehat{Y}_i(1)=\left\{\begin{array}{lll}
\frac{\sum_{j:W_j=1}^N Y_j K\left(\frac{p(X_j)-p(X_i)}{h}\right) }{\sum_{j:W_j=1}^N K\left(\frac{p(X_j)-p(X_i)}{h}\right)} & \text { if } & W_i=0, \\
Y_i & \text { if } & W_i=1.
\end{array}\right.
$$
Therefore, the kernel-matching estimator for the ATE is defined as
\begin{equation*}
\begin{split}
\hat{\tau}_{N}=&\frac{1}{N} \sum_{i=1}^N\left(\hat{Y}_i(1)-\hat{Y}_i(0)\right)\\
=&\frac{1}{N} \sum_{i=1}^N\left(2 W_i-1\right)\left(Y_i-\frac{\sum_{j:W_j=1-W_i}^N Y_j K\left(\frac{p(X_j)-p(X_i)}{h}\right)}{\sum_{j:W_j=1-W_i}^N K\left(\frac{p(X_j)-p(X_i)}{h}\right)}\right),
\end{split}
\end{equation*}
and the kernel-matching estimator for the ATT is defined as 
\begin{equation*}
\begin{split}
\hat{\tau}_{t}=\frac{1}{N_1}\sum_{i=1}^N W_i\left(Y_i-\frac{\sum_{j=1}^N (1-W_j) Y_j K\left(\frac{p(X_j)-p(X_i)}{h}\right)}{\sum_{j=1}^N (1-W_j) K\left(\frac{p(X_j)-p(X_i)}{h}\right)}\right), \end{split}    
\end{equation*}
where $N_1=\sum_{i=1}^N W_i$ is the number of treated units in the sample.

In observational studies, to estimate the propensity scores, we follow \cite{rosenbaum1983central} and use a generalized linear model $p(X) = F(X^\top \beta)$. The logit and probit functions are commonly used parametric link functions $F(\cdot)$. 

Let $\beta^*$ be the true value of the propensity score model parameter vector, such that $p(X) = F(X^\top \beta^*)$. The estimator based on matching the true propensity score is given by $\widehat{\tau}_N(\beta^*)$, and $\widehat{\tau}_N(\hat{\beta})$ is a kernel matching estimator based on the estimated propensity score $F(X^\top \hat{\beta})$, where $\hat{\beta}$ is the maximum likelihood estimator of $\beta$ based on the observed data $\{Y_i, W_i, X_i\}^N_{i=1}$, i.e., 
$$
\hat{\beta} = \arg \max_\beta \ln L\left(\beta \mid W_1, X_1, \ldots, W_N, X_N\right)
$$
with the log-likelihood function 
$$
\begin{aligned}
& \ln L\left(\beta \mid W_1, X_1, \ldots, W_N, X_N\right) \\
& \quad=\sum_{i=1}^N W_i \ln F\left(X_i^{T} \beta\right)+\left(1-W_i\right) \ln \left(1-F\left(X_i^{T} \beta\right)\right).
\end{aligned}
$$
Next, we define a type of matching estimator where the unknown parameter \( \beta \) is estimated using the maximum likelihood method. The kernel-matching estimators based on the  \(F( X^\top \hat{\beta}) \) for the ATE can be expressed as 
\begin{equation*}
 \begin{split}
\hat{\tau}_N(\hat{\beta}) &= \frac{1}{N} \sum_{i=1}^N (\hat{Y}_i(1) - \hat{Y}_i(0)) \\
&= \frac{1}{N} \sum_{i=1}^N \left(2 W_i - 1\right) \left(Y_i - \frac{\sum_{j: W_j = 1 - W_i} Y_j K\left( \frac{F(X_j^\top \hat{\beta}) - F(X_i^\top \hat{\beta})}{h} \right)}{\sum_{j: W_j = 1 - W_i} K\left( \frac{F(X_j^\top \hat{\beta}) - F(X_i^\top \hat{\beta})}{h} \right)}\right),
\end{split}   
\end{equation*}
and for ATT, the estimator can be expressed as
\begin{equation*}
\begin{split}
\hat{\tau}_{t,N}(\hat{\beta}) = \frac{1}{N_1} \sum_{i=1}^N W_i \left(Y_i - \frac{\sum_{j=1 }(1-W_j)  Y_j K\left(\frac{F(X_j^{\top} \hat{\beta}) - F(X_i^{\top} \hat{\beta})}{h}\right)}{\sum_{j=1 }(1-W_j) K\left(\frac{F(X_j^{\top} \hat{\beta}) - F(X_i^{\top} \hat{\beta})}{h}\right)}\right).
\end{split}    
\end{equation*}

\section{Asymptotic Properties of the Kernel-matching Estimators}
\label{sec:3} 

In this section, we establish that the kernel-matching estimators for the ATE and ATT are consistent and asymptotically normally distributed, with a convergence rate of $N^{1/2}$.

\begin{assumption}
The kernel function $K(\cdot)$ is symmetric and satisfies the following properties:
\begin{itemize} 
\item[{(1)}] $\int K(u)du=1$;
\item[{(2)}] $\int u^2K(u)du< \infty$;
\item[{(3)}] $\int K^2(u)du< \infty$;
\item[{(4)}] $\int |u|^3 K(u)du< \infty$.
\end{itemize}
\label{assum2}
\end{assumption}
Let $\mu_w(X)=E(Y(w) \mid X), \sigma_w^2(X)=\operatorname{Var}(Y(w) \mid X), \tau(X)=\mu_1(X)-\mu_0(X),  \pi = P(W=1).$ Similarly, define $\bar{\mu}_w(p)=E(Y(w) \mid p(X)=p), \bar{\sigma}_w^2(p)=\operatorname{Var}(Y(w) \mid  p(X)=p)$. Under Assumptions \ref{assum1} and \ref{assum2}, the following theorem shows that the estimators $\hat{\tau}$ and $\hat{\tau}_t$ are consistent estimators of $\tau$ and $\tau_t$, respectively, under mild conditions.

\begin{theorem}
Suppose that $\bar{\mu}_w(p)$ is three times continuously differentiable
and that the propensity score $p(X)$ is continuously distributed with probability density function $f(p)$ that is three times continuously differentiable, then under Assumptions \ref{assum1} and \ref{assum2}, as $N \rightarrow \infty$,  $Nh \rightarrow \infty$ and  $h \to 0$, we have 
\begin{itemize} 
\item[{(i)}]   
$$
\hat{\tau} - \tau \stackrel{p}{\longrightarrow} 0,
$$
\item[{(ii)}]  
$$  
\hat{\tau}_t-\tau_t \stackrel{p}{\longrightarrow} 0.
$$
\end{itemize}
\end{theorem}

Next, we elucidate the formal result for the asymptotic normality of the proposed estimators.
    
\begin{theorem}
Suppose that $\bar{\mu}_w(p)$ is three times continuously differentiable and that the propensity score $p(X)$ is continuously distributed with probability density function $f(p)$ that is three times continuously differentiable. Let
$$
b_w(r)=\frac{1}{2}\left\{\int t^2 K(t) \mathrm{d} t\right\}\left[\bar{\mu}_w^{\prime \prime}(r)+\frac{2 \bar{\mu}_w^{\prime}(r) f_{p(X) \mid W=w}^{\prime}(r)}{f_{p(X) \mid W=w}(r)}\right], \quad w \in\{0,1\} .
$$
and under Assumptions \ref{assum1} and \ref{assum2}, as $N \rightarrow \infty$, $h\to 0$, $Nh \rightarrow \infty$ and $Nh^4 = O(1)$, we have 
\begin{itemize} 
\item[{(i)}]   
$$
\sqrt{N}\left(\hat{\tau} - \tau -Bh^2\right)\xrightarrow{d}\mathcal{N}(0,\sigma^2),
$$
where $B=-\operatorname{Pr}(Z=1) E\left[b_0\{p(U)\} \mid Z=1\right]+\operatorname{Pr}(Z=0) E\left[b_1\{p(U)\} \mid Z=0\right]$, and the formula for $\sigma$ is provided in Appendix A;
\item[{(ii)}]
$$
\sqrt{N}\left(\hat{\tau}_t - \tau_t -B_th^2\right)\xrightarrow{d}\mathcal{N}(0,\sigma_t^2),
$$
where $B_t=-E\left[b_0\{p(U)\} \mid Z=1\right]$ and the formula for $\sigma_t$ is provided in Appendix A.
\end{itemize} 
\end{theorem}

The proofs of Theorem 1 and Theorem 2 are given in Appendix A.

\begin{corollary}
Suppose that $\bar{\mu}_w(p)$ is three times continuously differentiable, and the propensity score $p(X)$ is continuously distributed, with a continuous probability density function $f(p)$ which is three times continuously differentiable. Under Assumptions \ref{assum1} and \ref{assum2}, as $N \rightarrow \infty$, $h\to 0$, $Nh \rightarrow \infty$ and  $Nh^4\rightarrow 0$, the bias term will converge to zero. Thus,
\begin{itemize} 
\item[{(i)}]
$$
\sqrt{N}(\hat{\tau} - \tau )\xrightarrow{d}\mathcal{N}(0,\sigma^2);
$$
\item[{(ii)}]
$$
\sqrt{N}(\hat{\tau}_t - \tau_t) \xrightarrow{d}\mathcal{N}(0,\sigma_t^2), 
$$  
\end{itemize}
\end{corollary}

\begin{remark}
Nearest-neighbor (NN) matching and kernel matching can both be viewed as weighted-imputation methods. NN matching imputes each missing potential outcome using a fixed number of equally weighted matches, whereas kernel matching assigns weights according to distance, with the bandwidth determining the effective number of matches. Indeed, fixed-$M$ NN matching is a special case of kernel matching obtained by using a uniform kernel and setting the observation-specific bandwidth equal to the distance to the $M$th nearest neighbor. For both the ATE and ATT, fixed-$M$ matching introduces an additional nonnegative variance component of order $O(M^{-1})$. Consequently, as $M$ diverges, NN matching and kernel matching have the same first-order asymptotic variance, although their higher-order properties may differ \citep{abadie2016matching,hf}.
\end{remark}

In the following results, we will provide the large-sample properties of $\hat{\tau}_N(\hat{\beta})$ and $\hat{\tau}_{t,N}(\hat{\beta})$, including their  consistency in Theorem 4 and the 
asymptotic normality in Theorem 5. 

\begin{theorem}
Suppose that $\bar{\mu}_w(p)$ is three times continuously differentiable, and the propensity score $p(X)$ is distributed with a continuous probability density function $f(p)$ which is three times continuously differentiable. Under Assumptions \ref{assum1} and \ref{assum2}, as $N \rightarrow \infty$, $h\to 0$ and $Nh \rightarrow \infty$, we have
\begin{itemize}
\item[{(i)}]  
$$
\widehat{\tau}_N(\hat{\beta})-\tau\xrightarrow{p}0,
$$
\item[{(ii)}]  
$$  
\widehat{\tau}_{t,N}(\hat{\beta})-\tau_t\xrightarrow{p}0.
$$
\end{itemize}
\end{theorem}

\begin{theorem}
Suppose $\bar{\mu}_w(p)$ is three times continuously differentiable and bounded. Given that the propensity score $p(X)$ is continuously distributed, with continuous density, $f(p)$  has a continuous third derivative.
Under Assumptions \ref{assum1} and \ref{assum2}, as $N\rightarrow \infty$, $h\to 0$, $Nh \rightarrow \infty$ and $Nh^4=O(1)$, we have
\begin{itemize} 
\item[{(i)}]    
$$
\sqrt{N}\left\{\widehat{\tau}_N(\hat{\beta}) - \tau -Bh^2\right\}\xrightarrow{d}\mathcal{N}(0,\widetilde{\sigma}^2),
$$
where  $\widetilde{\sigma}^{2}
=
\sigma^{2}
-
\widetilde{T}(F)I_{\beta^*}^{-1}\widetilde{T}(F)^{\top}$, 
\item[{(ii)}] 
$$
\sqrt{N}\left\{\widehat{\tau}_{t,N}(\hat{\beta}) - \tau_t -B_th^2\right\}\xrightarrow{d}\mathcal{N}(0,\widetilde{\sigma}_t^2),
$$
where  $\widetilde\sigma_t^2
= \sigma_t^2
-c_t^\top I_{\beta^*}^{-1}c_t
+a_t^\top I_{\beta^*}^{-1}a_t$.
\end{itemize} 
\end{theorem}

The proof of Theorem 5 is presented in Appendix A.

\begin{corollary}
Suppose that $\bar{\mu}_w(p)$ is three times continuously differentiable, and the propensity score $p(X)$ is distributed with a continuous probability density function $f(p)$ which is three times continuously differentiable. Under Assumptions \ref{assum1} and \ref{assum2}, as $N \rightarrow \infty$, $h\to 0$, $Nh \rightarrow \infty$ and  $Nh^4\rightarrow 0$, the bias term will converge to zero. Thus, we have 
\begin{itemize} 
\item[{(i)}]    
$$
\sqrt{N}\left\{\widehat{\tau}_N(\hat{\beta}) - \tau \right\}\xrightarrow{d}\mathcal{N}(0,\widetilde{\sigma}^2),
$$
\item[{(ii)}]  
$$
\sqrt{N}\left\{\widehat{\tau}_{t,N}(\hat{\beta}) - \tau_t\right\}\xrightarrow{d}\mathcal{N}(0,\widetilde{\sigma}_t^2),
$$
\end{itemize} 
\end{corollary}

\begin{remark}
Since $I_{\beta^*}$ is positive definite, the term $\widetilde{T}(F)I_{\beta^*}^{-1}\widetilde{T}(F)^{\top}$ is non-negative. Therefore, when $\widetilde{T}(F) \neq 0$, the kernel-matching ATE estimator based on the estimated propensity score is asymptotically more efficient than the estimator based on the true propensity score. If the outcome regression functions depend on the covariates only through the propensity score, then $\widetilde{T}(F)=0$, and estimating the propensity score does not yield any first-order efficiency gain for the ATE estimator. 
In contrast, its effect on the ATT estimator has no definite sign. 
Accordingly, the effect of propensity score estimation on the asymptotic variance of the ATT estimator is not determined in general; it may decrease, remain unchanged, or increase depending on the sign of the term $c_t^{\top} I_{\beta^*}^{-1} c_t-a_t^{\top} I_{\beta^*}^{-1} a_t$.
Thus, unlike the ATE case, propensity score estimation does not uniformly improve the efficiency of the ATT estimator; its effect depends on the relative magnitudes of the score--influence covariance and treated-population composition terms. These conclusions parallel \cite{abadie2016matching}, but are derived here for kernel matching.

To assess the finite-sample implications of these theoretical results, Appendix B reports a simulation study comparing the kernel-matching ATE and ATT estimators based on the true and estimated propensity scores. The simulations evaluate their relative efficiency and provide numerical evidence consistent with the theoretical conclusions: estimating the propensity score improves or preserves the first-order efficiency of the ATE estimator, whereas its effect on the efficiency of the ATT estimator depends on the underlying data-generating process.
\end{remark}

\subsection{Comparison with semiparametric efficiency bounds}

Under the unconfoundedness and overlap assumptions, \cite{hahn1998role} derived the semiparametric efficiency bounds for the ATE and ATT estimators. Kernel matching based solely on the propensity score does not generally attain these bounds, because reducing the covariates to the propensity score may discard information relevant to the conditional outcome distributions. Nevertheless, we identify two settings in which the proposed estimators can achieve semiparametric efficiency.

First, consider the sufficient dimension reduction conditions
\[
\bar{\mu}_w\{p(X)\}=\mu_w(X)
\quad\text{and}\quad
\bar{\sigma}_w^2\{p(X)\}=\sigma_w^2(X),
\qquad w=0,1.
\]
These conditions hold when the conditional means and variances of the potential outcomes depend on the covariates only through the propensity score. In this case, the propensity score retains all information needed for efficient estimation, and the asymptotic variances of the kernel-matching estimators coincide with the corresponding semiparametric efficiency bounds for both the ATE and ATT. This result follows an argument similar to that of \cite{hf} and reinforces the observation of \cite{hahn1998role} that the main role of the propensity score in efficient estimation is to provide sufficient dimension reduction.

Second, we consider augmented, doubly robust kernel-matching estimators under a more general sufficient dimension reduction structure. Specifically, suppose there exists a fixed-dimensional representation \(R=X^{\top}\beta\), with \(\dim(R)\ll\dim(X)\), such that unconfoundedness holds conditional on \(R\), and both the propensity score and the conditional outcome regression functions depend on \(X\) only through \(R\). The unknown link functions are estimated nonparametrically by kernel smoothing, while the index parameters are estimated using sliced inverse regression. Under these conditions, the leading asymptotic representations of the doubly robust ATE and ATT estimators coincide with their respective efficient influence functions. Consequently, both estimators attain the semiparametric efficiency bounds. This approach also reduces the dimensionality burden associated with applying nonparametric smoothing directly to the full covariate vector.

Appendix C provides the efficiency bounds, formal conditions, asymptotic results, consistent variance estimators, and detailed proofs. It also presents simulation studies in which the estimation errors decrease with the sample size, the empirical coverage probabilities remain close to their nominal coverage rates, and the normalized standard deviations approach the square roots of the corresponding efficiency bounds. These findings provide finite-sample evidence consistent with the theoretical efficiency results.

\subsection{Consistent Variance Estimation}

In this subsection, we construct consistent estimators of the asymptotic variances of the ATE and ATT estimators to facilitate statistical inference. For the ATE estimator, a consistent variance estimator is given by
$$
\begin{gathered}
\widehat{\sigma}^2=\frac{1}{N} \sum_{i=1}^N \Bigg[\frac{W_i}{\hat{p}_i}\left\{Y_i-\widehat{\bar{\mu}}_1\left(\hat{p}_i\right)\right\}-\frac{1-W_i}{1-\hat{p}_i}\left\{Y_i-\widehat{\bar{\mu}}_0\left(\hat{p}_i\right)\right\} \\
+\widehat{\bar{\mu}}_1\left(\hat{p}_i\right)-\widehat{\bar{\mu}}_0\left(\hat{p}_i\right)-\widehat{\tau}\Bigg]^2 .
\end{gathered}
$$
Similarly, for the ATT estimator, a consistent estimator of its asymptotic variance is given by
\[
\begin{aligned}
\widehat{\sigma}_t^2
=
\frac{1}{N\widehat{\pi}^2}
\sum_{i=1}^N
\Bigg[
&
W_i
\left\{
Y_i-\widehat{\bar{\mu}}_0(\widehat p_i)
-\widehat{\tau}_t
\right\}^2
\\
&+
(1-W_i)
\frac{\widehat p_i^2}
     {(1-\widehat p_i)^2}
\left\{
Y_i-\widehat{\bar{\mu}}_0(\widehat p_i)
\right\}^2
\Bigg].
\end{aligned}
\]

Finally, we discuss consistent estimators of the asymptotic variances of the ATE and ATT estimators based on the estimated propensity score. A consistent estimator of $\widetilde{T}(F)$ is
$$
\widehat{\tilde{T}(F)}=-\frac{1}{N} \sum_{i=1}^N\left\{\frac{\widehat{C}_{0, i}^{\top}}{1-\hat{p}_i}+\frac{\widehat{C}_{1, i}^{\top}}{\hat{p}_i}\right\},
$$
where 
$$
\begin{aligned}
\widehat{C}_{w, i}= & \frac{\sum_{j \neq i} \mathbf{1}\left(W_j=w\right) K\left(\frac{\widehat{p}_j-\widehat{p}_i}{h}\right) f\left(X_j^{\top} \widehat{\beta}\right) X_j Y_j}{\sum_{j \neq i} \mathbf{1}\left(W_j=w\right) K\left(\frac{\widehat{p}_j-\widehat{p}_i}{h}\right)} \\
& -\left\{\frac{\sum_{j \neq i} \mathbf{1}\left(W_j=w\right) K\left(\frac{\widehat{p}_j-\widehat{p}_i}{h}\right) f\left(X_j^{\top} \widehat{\beta}\right) X_j}{\sum_{j \neq i} \mathbf{1}\left(W_j=w\right) K\left(\frac{\widehat{p}_j-\widehat{p}_i}{h}\right)}\right\} \\
& \times\left\{\frac{\sum_{j \neq i} \mathbf{1}\left(W_j=w\right) K\left(\frac{\widehat{p}_j-\widehat{p}_i}{h}\right) Y_j}{\sum_{j \neq i} \mathbf{1}\left(W_j=w\right) K\left(\frac{\widehat{p}_j-\widehat{p}_i}{h}\right)}\right\},
\end{aligned}
$$
Then, the Fisher information matrix is consistently estimated by
$$
\begin{aligned}
\hat{I}_{\beta^*}
&=\frac{1}{N} \sum_{i=1}^N \frac{f\left( \hat{\beta}\right)^2}{F\left( \hat{\beta}\right)\left(1-F\left( \hat{\beta}\right)\right)} X_i X_i^{\top}.    
\end{aligned}   
$$
A consistent estimator of the asymptotic variance when the propensity score is estimated is then
\begin{equation}
\widehat{\widetilde{\sigma}}^2= \hat{\sigma}^2-\widehat{\tilde{T}(F)} \hat{I}_{\beta^*}^{-1} \widehat{\tilde{T}(F)} ^{\top}.
\label{eq:estimated-ps-ate-variance}
\end{equation}
Define
$\widehat\omega_{ij}
=
\frac{\mathbf 1(j\ne i)K_{ij}}
{\sum_{\ell\ne i}K_{i\ell}}$,
$\widehat m_{\dot p,i}
=
\sum_{j\ne i}\widehat\omega_{ij}\widehat{\dot p}_j$,
$\widehat\pi
=
\frac1N\sum_{i=1}^N W_i.$
Using the conditional-expectation decompositions in Theorem 5, consistent
estimators of \(a_t\) and \(c_t\) are
\begin{align*}
\widehat a_t
&=
\frac{1}{N\widehat\pi}
\sum_{i=1}^N
\left[
\widehat C_{1,i}-\widehat C_{0,i}
+
\widehat m_{\dot p,i}
\left\{
\widehat{\bar\mu}_{1,i}
-\widehat{\bar\mu}_{0,i}
-\widehat\tau_t
\right\}
\right]\\
\mbox{and } \widehat c_t
&=
\frac{1}{N\widehat\pi}
\sum_{i=1}^N
\left[
\widehat C_{1,i}
+
\frac{\widehat p_i}{1-\widehat p_i}\widehat C_{0,i}
+
\widehat m_{\dot p,i}
\left\{
\widehat{\bar\mu}_{1,i}
-\widehat{\bar\mu}_{0,i}
-\widehat\tau_t
\right\}
\right],
\end{align*}
respectively. The consistent first-step-adjusted ATT variance estimator is therefore
\begin{equation}
\widehat{\widetilde{\sigma}}_t^2=\widehat{\sigma}_t^2-\hat{c}_t^{\top} \hat{I}_{\beta^*}^{-1} \hat{c}_t+\widehat{a}_t^{\top} \hat{I}_{\beta^*}^{-1} \widehat{a}_t .
\label{eq:estimated-ps-att-variance}
\end{equation}
Under the overlap, smoothness, moment, and bandwidth conditions stated in
Theorem 5, the estimators in
Eqs. \eqref{eq:estimated-ps-ate-variance} and
\eqref{eq:estimated-ps-att-variance} are consistent. The corresponding
standard-error estimators are
$\widehat{\operatorname{se}}(\widehat\tau)
=
\sqrt{\frac{\widehat{\widetilde\sigma}^2}{N}}$ and 
$\widehat{\operatorname{se}}(\widehat\tau_t)
=
\sqrt{\frac{\widehat{\widetilde\sigma}_t^2}{N}},
$ for the estimators in Eqs. \eqref{eq:estimated-ps-ate-variance} and
\eqref{eq:estimated-ps-att-variance}, respectively.

\color{black}
\section{Monte Carlo Simulation Studies}
\label{sec:4}

\subsection{Simulation settings}
\label{sec:4.1}
In this section, we conduct Monte Carlo simulation studies to evaluate the performance of the proposed estimators and compare them with existing estimators. The simulation results confirm the effectiveness of the kernel matching estimators based on the estimated propensity score in estimating ATE and ATT. For comparative purposes, we consider the following existing ATE and ATT estimation methods: 
\begin{itemize} 
\item Abadie-Imbens matching estimators based on the original covariates \citep{abadie2006large} (denoted as ``Covariate"). 
\item Abadie-Imbens matching estimators based on the true propensity score \citep{abadie2016matching} (denoted as ``True PS"). 
\item Abadie-Imbens matching estimators based on the estimated propensity score \citep{abadie2016matching} (denoted as ``Estimated PS"). 
\item Normalized inverse probability weighting (IPW) \citep{hirano2003efficient, imbens2004nonparametric} estimators for ATE and ATT given by 
$$
\hat{\tau}_{\mathrm{IPW}}^{\mathrm{ATE}}=\sum_{i=1}^N \frac{W_i Y_i}{\hat{p}\left(X_i\right)} \left( \sum_{i=1}^N \frac{W_i}{\hat{p}\left(X_i\right)} \right)^{-1} - \sum_{i=1}^N \frac{\left(1-W_i\right) Y_i}{1-\hat{p}\left(X_i\right)} \left(\sum_{i=1}^N \frac{\left(1-W_i\right)}{1-\hat{p}\left(X_i\right)}\right)^{-1} 
$$
and
$$
\hat{\tau}_{\mathrm{IPW}}^{\mathrm{ATT}} =   \frac{1}{N_1} \sum_{i: W_i=1} Y_i  -  \sum_{i: W_i=0} Y_i \cdot \frac{\hat{p}(X_i)}{1-\hat{p}(X_i)} \left(\sum_{i: W_i=0} \frac{\hat{p}(X_i)}{1-\hat{p}(X_i)} \right)^{-1},
$$
respectively (denoted as ``IPW").
\item Doubly robust estimators  
\citep{robins1994estimation,bang2005doubly} for ATE and ATT given by 
\begin{eqnarray*}
\hat{\tau}_{\mathrm{DR}}^{\mathrm{ATE}} & = & \frac{1}{N} \sum_{i=1}^N\left\{\frac{W_i Y_i}{\hat{p}\left(X_i\right)}-\frac{W_i-\hat{p}\left(X_i\right)}{\hat{p}\left(X_i\right)} \hat{\mu}\left(1, X_i\right) \right. \\ 
& & \left. \qquad - \frac{\left(1-W_i\right) Y_i}{1-\hat{p}\left(X_i\right)}-\frac{W_i-\hat{p}\left(X_i\right)}{1-\hat{p}\left(X_i\right)} \hat{\mu}\left(0, X_i\right)\right\}
\end{eqnarray*}
and
\begin{eqnarray*}
\hat{\tau}_{\mathrm{DR}}^{\mathrm{ATT}}=\frac{1}{N_{1}} \sum_{i=1}^N\left[ W_i Y_i-\frac{\left(1-W_i\right) Y_i\hat{p}\left(X_i\right)+\hat{\mu}\left(0, X_i\right)\left(W_i-\hat{p}\left(X_i\right)\right)}{1-\hat{p}\left(X_i\right)}\right],
\end{eqnarray*}
respectively (denoted as ``DR").
\item  The kernel-matching estimators for ATE and ATT are presented in Section \ref{sec:2}  (denoted as ``Proposed").  Since the performance of kernel-based estimators depends on the choice of bandwidth and kernel function, selecting appropriate values for these parameters is essential. We discuss the selection criteria in detail in the following subsection.

\end{itemize}
To simulate the control and treatment units with covariates, we design the following five settings.
The data generation processes for Settings I and II follow the approach of \cite{abadie2016matching}, while those for Settings III and IV are based on the method of \cite{10.1162/003465304323023697} with slight modifications. Similarly, the process for Setting V is adapted from the approach of \cite{luo2020matching}, also with slight modifications.

\begin{itemize} 
\item \textbf{Setting I:} 
\begin{itemize} 
\item Two independent covariates, \(X_1\) and \(X_2\), which follow the uniform distribution in the interval \([-1/2, 1/2]\). 
\item The potential outcomes are generated based on the following models  
$$Y(0)=3X_1-3X_2+\epsilon_0$$
and 
$$Y(1)=5+5X_1+X_2+\epsilon_1,$$
where $\epsilon_0$ and $\epsilon_1$ are random errors that follow the standard normal distribution independent of $(W, X_1, X_2)$. 
\item The treatment variable, $W$, is related to $(X_1, X_2)$ through the propensity score with a logistic function
$$
\Pr(W=1|X_1=x_1,X_2=x_2)=\frac{\exp(x_1+2x_2)}{1+\exp(x_1+2x_2)}.
$$
\item The ATE and ATT estimators use a single match ($M = 1$).
\end{itemize} 
\item \textbf{Setting II:} 
\begin{itemize} 
\item Two independent covariates, \(X_1\) and \(X_2\), follow the uniform distribution in the interval \([-1/2, 1/2]\).
\item The potential outcomes are generated based on the following models  
$$
Y(0)=10 X_1+\epsilon_0
$$
and
$$
Y(1)=5 - 10X_1+\epsilon_1,
$$
where $\epsilon_0$ and $\epsilon_1$ are random errors that follow the standard normal distribution independent of $(W, X_1, X_2)$. 
\item The treatment variable, $W$, is related to $X_2$ through the propensity score with a logistic function
$$
\Pr(W=1|X_1=x_1,X_2=x_2)=\frac{\exp(2x_2)}{1+\exp(2x_2)}.
$$
\item The ATE and ATT estimators use a single match ($M = 1$).
\end{itemize}
\item \textbf{Setting III:} 
\begin{itemize} 
\item Four covariates, the vector $X = (X_1, X_2, X_3, X_4)$ follows a four-dimensional normal distribution $\mathcal{N}\left(0, \bSigma \right)$, where the variance-covariance $\bSigma$ is given by 
\begin{eqnarray*}
\bSigma = \frac{1}{3}
\begin{pmatrix}
1 & -1 & 0 & 0 \\ 
-1 & 2 & 0 & 0 \\
0 & 0 & 1 & -1 \\ 
0 & 0 & -1 & 2 \\
\end{pmatrix},
\end{eqnarray*}
i.e., $X_1$ and $X_2$ are correlated, $X_3$ and $X_4$ are correlated, and $(X_{1}, X_{2})$ and $(X_{3}, X_{4})$ are independent.
\item The potential outcomes are generated based on the following models
$$
Y=0.4+0.25\sin(8(X^\top \btheta)-5)+0.4\exp(-16(4(X^\top\btheta)-2.5)^2)+W+\epsilon,
$$  
where $\btheta = (1,1,1,1)^\top$, $\epsilon$ is a random error that follows the standard normal distribution independent of $X$.
\item The treatment variable, $W$, is related to $X = (X_1,X_2,X_3,X_4)$ through the propensity score with a logistic function
\begin{eqnarray*}
\Pr(W=1&|&X_1=x_1,X_2=x_2,X_3=x_3, X_{4} = x_{4}) \\
& = & \frac{\exp(x_1-3x_2+2x_3+x_4)}{1+\exp(x_1-3x_2+2x_3+x_4)}.
\end{eqnarray*}
\item The ATE and ATT estimators use a single match ($M = 1$).
\end{itemize} 
\item \textbf{Setting IV:} 
\begin{itemize}
\item Four covariates, the vector $X = (X_1, X_2, X_3, X_4)$ follows a four-dimensional normal distribution $\mathcal{N}\left(0, \bSigma \right)$, where the variance-covariance $\bSigma$ is given by 
\begin{eqnarray*}
\bSigma = \frac{1}{3}
\begin{pmatrix}
1 & -1 & 0 & 0 \\ 
-1 & 2 & 0 & 0 \\
0 & 0 & 1 & -1 \\ 
0 & 0 & -1 & 2 \\
\end{pmatrix}.
\end{eqnarray*}
\item The potential outcomes are generated based on the following models
$$
Y=0.15 + 0.7 \left[
\frac{\exp(\sqrt{2}(X^\top \btheta))}{ 1 + \exp(\sqrt{2}(X^\top \btheta))} \right]+ W + \epsilon,
$$  
where $\btheta = (1, 1, 1, 1)^\top$ and $\epsilon$ is a random error that follows the standard normal distribution independent of $X$.  
\item The treatment variable, $W$, is related to $X = (X_1,X_2,X_3,X_4)$ through the propensity score with a logistic function
\begin{eqnarray*}
\Pr(W=1&|&X_1=x_1,X_2=x_2,X_3=x_3,X_4=x_4)\\
& = & \frac{\exp(x_1+x_2+x_3+x_4)}{1+\exp(x_1+x_2+x_3+x_4)}.
\end{eqnarray*}
\item The ATE and ATT estimators use a single match ($M = 1$).
\end{itemize}
\item \textbf{Setting V:} 
\begin{itemize}

\item Three independent covariates \( X_1 \sim \mathcal{N}(0, 1) \), \( X_2 \sim \text{Bernoulli}(0.5) \), and \( X_3 \sim \mathcal{N}(0, 1) \).

\item The potential outcomes are generated based on the following models
$$
Y(0) = \sin(1 + X_1 + 0.5 X_2) + \epsilon_0
$$ 
and
$$
Y(1) = \sin(2 +X_1 + 0.5  X_2)  + \epsilon_1,
 $$  
where $\epsilon_0$ and $\epsilon_1$ are random errors that follow the standard normal distribution independent of $X$.   
\item The treatment variable, $W$, is related to $X = (X_1,X_2,X_3)$ through the propensity score with a logistic function
\begin{eqnarray*}
\Pr(W=1&|&X_1=x_1,X_2=x_2,X_3=x_3)\\
& = & \frac{\exp(2x_1 + x_2 - 3x_3 + 3)}{1+\exp(2x_1 + x_2 - 3x_3 + 3)}.
\end{eqnarray*}
\item The ATE and ATT estimators use a single match ($M = 1$).
\end{itemize}
\end{itemize} 

We generated datasets with sample sizes of $N = 200, 500$, and $1000$, and conducted $1000$ Monte Carlo simulations for each scenario to estimate causal effects.

\subsection{Selection of bandwidth and kernel functions in kernel matching}
\label{sec:4.2}

The efficacy of kernel matching estimators is contingent upon the appropriate selection of the kernel function and bandwidth parameter  ($h$). To rigorously evaluate the impact of these factors, we performed a comprehensive analysis of treatment effects using our proposed estimators across a range of bandwidth values, specifically $h = 0.01$ to $0.09$, with sample sizes of $N = 200, 500$, and $1000$. We consider two commonly used kernel functions: the Gaussian and Epanechnikov kernels. The analysis was conducted within the experimental framework of Settings I--V in Section \ref{sec:4.1}, and the results are presented in Figures \ref{fig:ATT1}--\ref{fig:ATT5} of Appendix D.

For the choice of kernel function, Figures \ref{fig:ATT1}--\ref{fig:ATT5} indicate that the relative performance of the Gaussian and Epanechnikov kernels depends on the simulation setting, sample size, and bandwidth. Neither kernel uniformly dominates the other in terms of bias and RMSE across all configurations. Under the bandwidth grid considered in our simulations, the Gaussian kernel often provides stable finite-sample performance and is therefore used in the subsequent experiments. This choice should be viewed as an empirical implementation choice rather than evidence of the
general theoretical superiority of the Gaussian kernel. In particular, the infinite support of the Gaussian kernel does not, by itself, eliminate the boundary bias of the local-constant Nadaraya--Watson estimator. Furthermore, the Gaussian kernel exhibits greater stability in numerical computation and gradient optimization, particularly in finite samples, where its advantages are more pronounced.

Bandwidth selection involves a trade-off between bias and variance of the estimators. As \cite{ullah1999nonparametric} pointed out, wider bandwidths result in smoother density estimates, which reduce variance at the cost of increased bias. Our simulation results confirm this, showing that estimator bias increases and RMSE decreases with larger bandwidths ($h$). This underscores the importance of bandwidth selection in achieving asymptotic properties. Theorems 2 and 5 establish that asymptotic normality requires $N h^4 \to 0$, or equivalently $h = o(N^{-1/4})$, as $N \to \infty$. In the simulation study, we use bandwidths approximately proportional to $N^{-1/2}$, which satisfy this requirement. This results in bandwidth values of $h = 0.07, 0.05$, and 0.03 for sample sizes of $N = 200, 500$, and 1000, respectively. 
Moreover, under the bandwidth grid considered in our simulations, the Gaussian kernel generally exhibited stable finite-sample performance and was therefore adopted in the subsequent experiments.

\subsection{Performance of estimators}
\label{sec:4.3}

Based on the results in Section \ref{sec:4.2}, we adopted the Gaussian kernel function and bandwidth values of $h = 0.07, 0.05$, and 0.03 for sample sizes of $N = 200, 500$, and 1000, respectively, in the simulation study with Settings I--V. We compare the point estimation performance of the six ATE and ATT estimators (Covariate, True PS, Estimated PS, Proposed, IPW, and DR) in terms of the simulated biases (Bias), standard deviations (SD), and root mean squared errors (RMSE). For comparative purposes, we also compute the 95\% confidence intervals based on different estimators. We estimate the standard errors (SEs) using $\widehat{\operatorname{se}}(\widehat\tau)$ and $\widehat{\operatorname{se}}(\widehat\tau_t)$.  We compare the interval estimation of ATE and ATT of the six estimators in terms of the simulated average widths (AW) and the coverage probabilities (CP). The simulation results are reported for the six estimators in Appendix E under Settings I--V in Tables \ref{tab:S1ATE}--\ref{tab:S5ATE} for ATE estimators and in Tables \ref{tab:S1ATT}--\ref{tab:S5ATT} for ATT estimators.  

From Tables \ref{tab:S1ATE}--\ref{tab:S5ATT}, we observe that the biases and RMSEs of the proposed estimator decrease as the sample size increases, which is consistent with the theoretical asymptotic results discussed in Section \ref{sec:3}. 
For point estimation of both ATE and ATT, the proposed kernel matching estimator demonstrates comparable bias and RMSE performance across most settings, though it does not achieve the smallest bias and RMSE in every scenario. From Tables \ref{tab:S1ATE}--\ref{tab:S2ATE}, the proposed ATE estimator generally has a lower RMSE than the matching estimators based on the true or estimated propensity score. Its RMSE is comparable to that of covariate matching, although covariate matching performs slightly better in some configurations, particularly in Setting II. From Tables \ref{tab:S3ATE}--\ref{tab:S4ATE}, we observe that the proposed ATE estimator has lower bias than the covariate-based matching estimator in Settings III and IV. This consistent performance highlights the effectiveness of the proposed estimator across varied settings, even when it does not uniformly outperform all competitors.

In terms of interval estimation, the proposed estimator generally achieves reasonable coverage probabilities, although some undercoverage occurs for the ATT in smaller samples, suggesting strong reliability of the proposed method. While the IPW estimator offers good point estimation, it struggles with interval estimation accuracy, failing to maintain 95\% coverage in Settings IV and V. Similarly, the DR estimator performs well for point estimation in Settings I–IV but falters in Setting V. The proposed estimator’s interval estimation remains stable, often surpassing IPW and DR in coverage accuracy, particularly in settings with extreme true propensity scores. Based on these simulation results, DR, IPW, and the proposed estimators are generally recommended; however, the proposed estimator is competitive, particularly in terms of interval estimation in Settings IV and V.

\subsection{Model misspecification}
\label{sec:4.4}

To assess the efficacy of our estimator under model misspecification, we conducted a Monte Carlo simulation study based on designs from \cite{busso2014new}, comparing its performance to alternative matching estimators.
This comparison was performed under conditions of both correctly and incorrectly specified outcome and selection equations. Although \cite{busso2014new} offered a comprehensive account of the data generation process, we provide a concise overview, with the foundational framework presented in the following equations:
\begin{eqnarray}
Y_i(0) & = & m\left(Z_i\right)+ \epsilon, \label{eq1} \\
T_i^* & = & Z_i-U_i, \label{eq2}
\end{eqnarray}
where $Z_i$ is a function of the covariates and $U_i$ follows a standard logistic distribution. The error term $\epsilon$ follows the standard normal distribution and is independent of $X_i$. 

The design adopts a linear model for both the outcome and selection equations. The covariate vector  $\left(X_1, X_2, X_3, X_4\right)^\top$ 
follows a multivariate normal distribution with mean zero, and the covariance matrix $\bSigma$, which is a block diagonal matrix 
\begin{eqnarray*}
\bSigma = \frac{1}{3}
\begin{pmatrix}
1 & -1 & 0 & 0 \\ 
-1 & 2 & 0 & 0 \\
0 & 0 & 1 & -1 \\ 
0 & 0 & -1 & 2 \\
\end{pmatrix},
\end{eqnarray*}
This configuration allows for correlation only between $X_1$ and $X_2$, and between $X_3$ and $X_4$. The model is built upon the structure described in Eqs. \eqref{eq1} and \eqref{eq2}. In this context, the latent treatment variable $T^*$ is generated using Eq. \eqref{eq2}. A linear function for $m(\cdot)$ is applied to generate $Y_i(0)$  using Eq. \eqref{eq1}. To define the treatment effect, a constant treatment effect of $1$ is assumed, so that $Y_i(1)=T_i+Y_i(0)$, where $T_{i} = \mathbb{I}\left(T_{i}^*>0\right)$, and $\mathbb{I}(\cdot)$ denotes the indicator function, which equals $1$ if the argument is true, and $0$ otherwise.

Based on this design, we examine scenarios involving both correct and misspecified propensity scores, with $N = 500$ observations. 
We consider the proposed estimator with the Gaussian kernel function with a bandwidth $h=0.04$, nearest-neighbor matching estimators with  
covariate matching (NN-Covariates) and propensity score matching (NN-PS) with different prespecified numbers of neighbors $k = 1, 4, 16,$ and 64, and the IPW estimator. 
The simulated absolute biases and variances based on $1000$ replications are presented in Table \ref{tab:DI}. 
From Table \ref{tab:DI}, we observe that our proposed estimator yields competitive absolute bias compared to nearest-neighbor matching estimators. The proposed estimator has substantially lower absolute bias than covariate matching and slightly lower absolute bias than the propensity-score nearest-neighbor estimators. IPW has the smallest absolute bias in both cases. The variance of the proposed estimator is comparable to that of IPW and the propensity-score matching estimators using multiple neighbors. This performance is sustained even when the propensity score is misspecified.  The key advantage of the proposed estimator, shared with IPW, is its ability to perform data-driven matching without requiring the pre-selection of a fixed number of matches.
The kernel matching approach presented in this paper offers researchers a data-adaptive alternative, demonstrating competitive performance relative to nearest-neighbor matching and IPW, particularly in scenarios with complex treatment assignment or misspecified propensity scores.


\setcounter{table}{10}

\begin{table}[]\centering
\caption{Simulated absolute biases and variances for the proposed estimator, the nearest-neighbor matching estimator, and the IPW estimator under correctly and incorrectly specified propensity scores.}
\label{tab:DI}
\begin{tabular}{ccc}
\hline
Method                      & \begin{tabular}[c]{@{}c@{}}Linear\\ (Correctly specified PS)\end{tabular} & \begin{tabular}[c]{@{}c@{}}Linear\\ +Interactions\\ (Misspecified PS)\end{tabular} \\ \hline
Absolute Bias$ \times 1000$ &                                                      &                                                                      \\ \hline 
Proposed                    & 10.2                                                 & 8.5                                                                  \\ \hline 
NN-Covariates                  &                                                      &                                                                      \\ \cline{1-1}
$k=1$                            & 129.0                                                & 129.0                                                                \\
$k=4$                            & 192.7                                                & 192.7                                                                \\
$k=16$                           & 314.8                                                & 314.8                                                                \\
$k=64$                           & 485.0                                                & 485.0                                                                \\ \hline 
NN-PS                          &                                                      &                                                                      \\ \cline{1-1}
$k=1$                            & 12.8                                                 & 9.7                                                                  \\
$k=4$                            & 16.8                                                 & 13.2                                                                 \\
$k=16$                           & 43.1                                                 & 42.6                                                                 \\
$k=64$                           & 167.5                                                & 165.7                                                                \\ \hline 
IPW                            & 4.9                                                  & 4.5                                                                  \\ \hline \hline  
Variance$ \times n$          &                                                      &                                                                      \\ \hline 
Proposed                       & 6.5                                                  & 6.9                                                                  \\ \hline 
NN-Covariates                  &                                                      &                                                                      \\ \cline{1-1}
$k=1$                            & 8.5                                                  & 8.5                                                                  \\
$k=4$                            & 5.7                                                  & 5.7                                                                  \\
$k=16$                           & 5.6                                                  & 5.6                                                                  \\
$k=64$                           & 5.9                                                  & 5.9                                                                  \\ \hline 
NN-PS                          &                                                      &                                                                      \\ \cline{1-1}
$k=1$                            & 9.3                                                  & 9.8                                                                  \\
$k=4$                            & 6.6                                                  & 6.5                                                                  \\
$k=16$                           & 5.8                                                  & 6.1                                                                  \\
$k=64$                           & 5.5                                                  & 5.9                                                                  \\ \hline 
IPW                            & 6.6                                                  & 6.6                                                                  \\ \hline
\end{tabular}
\end{table}

\section{Practical Data Analysis}
\label{sec:5} 
In this section, we apply the proposed estimation procedures to analyze data from the National Supported Work (NSW) Demonstration.

\subsection{Analysis of data from the NSW Demonstration job-training program}

The NSW program was funded by both federal and private sources in the mid-1970s. This initiative provided 6--18 months of work experience for individuals facing economic and social challenges before enrollment. Our analysis demonstrates the impact of the NSW Demonstration labor training program on the earnings of these individuals. To evaluate the effectiveness of the proposed estimator, we estimate the ATT using two control groups and one treated group. We use two datasets: one drawn from the experimental sample in \cite{lalonde1986evaluating} and the other from Westat's Matched Current Population Survey–Social Security Administration File (CPS-3) \citep{DehejiaWahba1999} \footnote{The data can be obtained from the website at \href{https://users.nber.org/~rdehejia/nswdata2.html}{https://users.nber.org/$\sim$rdehejia/nswdata2.html}.}.

In this dataset, the outcome variable (Re78) represents each individual’s real earnings in 1978, and the treatment variable (Treat) indicates whether the individual enrolled in the labor training program. The ten potential confounding variables include age (Age), years of schooling (Educ), race indicators for Black (Black) and Hispanic (Hispanic), marital status (Married), high school diploma status (Nodegr), real earnings in 1974 (Re74) and 1975 (Re75), and indicators for zero earnings in 1974 (U74) and 1975 (U75). This subset comprises 185 individuals in the treated group, 260 individuals in the control group reported by \cite{lalonde1986evaluating}, and 429 individuals from the CPS-3 control group. Table \ref{tab:summary} presents the summary statistics for the NSW dataset and the $p$-values of the $t$-tests  for comparing the means between the treated group and the two control groups from the two samples.
Table \ref{tab:summary} shows that the treated and control groups are well-balanced in terms of mean values for most variables, with the exception of the variable Nodegr. For the nine remaining covariates, we cannot reject the null hypothesis that the means are equal, indicating similar averages across treated and control groups. However, when comparing treated units from the experimental data with control units from the non-experimental CPS-3 data, significant mean differences appear in eight out of the ten covariates. The only covariates for which the null hypothesis of equal means cannot be rejected are Hispanic and education.
\begin{table}[ht]
    \centering
    \caption{Summary statistics of the variables in the NSW dataset.}
    \label{tab:summary}
    
    \begin{tabular}{l*{6}{c}c}
        \toprule
        & \multicolumn{2}{c}{Experimental Data} & \multicolumn{1}{c}{Non-experimental CPS-3} && \multicolumn{2}{c}{$p$-value of two-sample $t$-test} \\
        \cmidrule(r){2-3} \cmidrule(r){4-5} \cmidrule(r){6-7}
        Variable & \multicolumn{1}{c}{Treated (185)} & \multicolumn{1}{c}{Control (260)} & \multicolumn{1}{c}{Control (429)} & & Treated/Control & Treated/Control \\
                 & Mean (SD) & Mean (SD) & Mean (SD) & & Experiments & Control CPS-3 \\
        \midrule
        Age      & 25.82 (7.16) & 25.1 (7.10)   & 28.03 (10.79) & & 0.27 & $< 0.001$ \\
        Educ     & 10.35 (2.01) & 10.1 (1.60)   & 10.24 (2.86)  & & 0.15 & 0.58 \\
        Black    &  0.84 (0.36) & 0.83 (0.38)  & 0.20 (0.40)   & & 0.65 & $< 0.001$ \\
        Hispanic & 0.06 (0.24) & 0.11 (0.31)  & 0.14 (0.35)   & & 0.06 & 0.13 \\
        Married  & 0.19 (0.39) & 0.15 (0.36)  & 0.51 (0.50)   & & 0.33 & $< 0.001$ \\
        Nodegr   & 0.71 (0.46) & 0.83 (0.37)   & 0.60 (0.49)   &  & $< 0.001$   & 0.01   \\
        Re74     & 2.10 (4.89) & 2.11 (5.69)  & 5.62 (6.79)   & & 0.98 & $< 0.001$ \\
        Re75     & 1.53 (3.22) & 1.27 (3.10)  & 2.47 (3.29)   & & 0.39 & $< 0.001$ \\
        U74      & 0.71 (0.46) & 0.75 (0.43)  & 0.26 (0.44)   & & 0.33 & $< 0.001$ \\
        U75      & 0.60 (0.49) & 0.68 (0.47)  & 0.31 (0.46)   & & 0.07 & $< 0.001$ \\
        \bottomrule
    \end{tabular}
    \footnotesize{\scriptsize Earnings are expressed in thousands of 1978 dollars. 
    }
\end{table}

\begin{table}[ht]
\centering
\caption{Estimates of ATT and their standard errors and corresponding 95\% confidence intervals based on different methods for the NSW dataset.}
\label{tab:estNSW}
\begin{tabular}{ccccc}
\hline
\textbf{}    & \multicolumn{2}{c}{\textbf{Experimental Group}} & \multicolumn{2}{c}{\textbf{CPS-3}}  \\
             & Estimate (SE)          & 95\% C.I               & Estimate (SE)   & 95\% C.I          \\ \hline
Proposed     & 1858.2 (669.8)        & (545.4, 3171.0)        & 848.2 {\color{red}(676.5)}   & {\color{red}(-477.7, 2174.1)}  \\
Covariates   & 1686.1 (866.4)         & (-12.0, 3384.2)        & -243.5 (1319.3) & (-2829.3, 2342.3) \\
Estimated PS & 2138.6 (797.8)         & (575.0, 3702.3)        & -944.9 (1383.7) & (-3656.9, 1767.1) \\
IPW          & 1754.6 (695.5)         & (391.4, 3117.8)        & 1003.4 (698.6)  & (-365.8, 2372.6)  \\
DR           & 1718.9 (751.1)         & (246.8, 3191.0)        & 836.0 (965.3)   & (-1056.0, 2728.0) \\ \hline
\end{tabular}
\end{table}

For the experimental data, the proposed kernel matching estimator with the Gaussian kernel function and bandwidth of $h = 0.05$ is used. For the non-experimental CPS-3 data, the proposed kernel matching estimator with the Gaussian kernel function and bandwidth of $h = 0.04$ is used.
In Table \ref{tab:estNSW}, we present the estimates of ATT using the proposed kernel matching estimator (``Proposed"), the matching estimator based on covariates (``Covariates"), the matching estimator based on estimated propensity scores with $M = 1$ (``Estimated PS"), and the IPW and DR estimators. 
For the experimental group, the ATT estimate based on the proposed kernel matching estimator is US\$1858.2, the closest to the benchmark value of US\$1794 \citep{DehejiaWahba1999} among the estimation methods considered here. Moreover, the ATT estimate based on the proposed kernel matching estimator has the smallest standard error among the estimation methods considered here. The proposed kernel matching estimator, the IPW, and the DR estimators give ATT estimates and confidence intervals within a similar range. For the non-experimental control group (CPS-3), the matching estimators based on covariates and estimated propensity scores produce negative estimates with large standard errors, which do not agree with the results obtained by \cite{DehejiaWahba1999}. In contrast, the proposed kernel matching estimator, along with IPW and DR methods, yields ATT estimates within a similar range. 
These findings suggest that the labor training program significantly increased individual earnings in 1978 and are consistent with the conclusions drawn from the experimental study conducted by \cite{DehejiaWahba1999}. 

\subsection{Assessing performance of estimators under bad and good overlap conditions}

Based on the data from the NSW Demonstration job-training program and  the Panel Study of Income Dynamics
(PSID) used by \cite{DehejiaWahba1999}, in this subsection, we present a Monte Carlo simulation study designed to compare our proposed estimator with nearest-neighbor matching (using covariates and propensity scores) with prespecified numbers of neighbors $k = 1, 4$, and 16, and IPW. The comparison is conducted under both bad and good overlap conditions, where overlap means the degree of similarity in propensity scores between treated and control units. Good overlap indicates substantial overlap in the propensity score distributions, facilitating more reliable estimation. Conversely, bad overlap, characterized by highly separated distributions, results in a scarcity of comparable controls for treated units, potentially inflating estimation bias and variability.

Following the Monte Carlo design of \cite{diaz2015matching}, this study compares an experimental African American subsample ($n = 156$) with a control group from the Panel Study of Income Dynamics (PSID) dataset, as described by Dehejia and Wahba (1999) ($n = 624$). Covariates include age, education, marital status, 1974 and 1975 earnings, unemployment status, and a high school dropout indicator. To capture nonlinearities, interaction terms between 1974 and 1975 unemployment and earnings variables, as well as squared earnings terms, are incorporated. We define $X_{i}$  as the covariate set excluding squared and interaction terms, and $Z_{i}$ as the full set, including these terms. The following equations outline the main framework of the data generation process:
\begin{eqnarray}
 Y_i(0) & = & \delta_0^{\prime} Z_i+\epsilon_{0 i}, \label{eq3}\\ 
 T_i^* & = & \alpha+\beta Z_i-U_i, \label{eq4} 
 \end{eqnarray}
where $Z_i$ is a function of the covariates (described previously) and $U_i$ follows the standard logistic distribution. 
Then,  $T^*$ is generated using Eq. \eqref{eq4}, where the coefficients $\alpha$ and $\beta$ are replaced with the corresponding estimates from a logistic regression on the original NSW sample. Instead of $\delta_0$, the coefficients from a regression of $Y_i(0)$ on $Z_i$ based on the control observations in the NSW sample are used. Similarly, $Y_i(1)$ is constructed by the regression model of $Y_i(1)$ on $Z_i$ using the treated units from the NSW sample. 

We generate 5,000 random samples with this setting, each of sample size $N=400$.  Based on the selection of the kernel function and bandwidth discussed in Section \ref{sec:4.2}, 
we evaluated our proposed ATT estimator with the Gaussian kernel function and bandwidth of $h=0.05$. Here, the benchmark for ATT is $\$2334$ \citep{busso2014new}. Building upon \cite{busso2014new}, 
we also examined estimator performance under good overlap conditions, in addition to the previously discussed bad overlap conditions. 
To achieve good overlap, the parameters in the treatment selection equation are scaled down by a factor of 5, thereby increasing the influence of the random component $\left(U_i\right)$  
and increasing the treatment assignment variability, which leads to improved overlap in the data.

Table \ref{tab:CI} presents the simulated absolute biases and variances for different estimators under both bad and good overlap conditions. 
Table \ref{tab:CI} shows that no estimator uniformly dominates the others. The proposed estimator provides a relatively stable compromise between bias and variance across the two overlap conditions. Some nearest-neighbor configurations and IPW achieve smaller bias or variance in particular settings, but their performance is more sensitive to the degree of overlap and, for nearest-neighbor matching, to the choice of $k$.
As previously discussed, nearest-neighbor estimators face a trade-off between bias and variance, where increasing the number of matches reduces variance but elevates bias.

Under good overlap, the qualitative findings are similar to those observed under bad overlap. Both our proposed estimator and IPW exhibit comparable performance in terms of bias and variance. Nearest-neighbor matching based on covariate matching achieves optimal performance with 4 neighbors, but its bias substantially increases with 16 neighbors. Similarly, nearest-neighbor matching based on propensity score achieves the smallest bias with 4 neighbors, but experiences a significant bias increase with 16 neighbors.

These simulation results highlight the advantages of the proposed estimator, which maintains robust performance across varying overlap conditions. First, its smoothing property, achieved through weighted averaging of neighboring observations, ensures stable estimation, reducing variance in favorable overlap scenarios and mitigating instability in regions with limited overlap. Second, kernel weighting demonstrates greater robustness to extreme weights compared to alternative methods, which can suffer from high variance in bad overlap conditions. Finally, kernel matching efficiently utilizes the entire control group by incorporating all available observations with appropriate weights, thereby avoiding the information loss inherent in nearest-neighbor matching. These characteristics may contribute to its relatively stable performance across the two overlap conditions.

\begin{table}[H]
\centering
\caption{Simulated absolute biases and variances for different estimators under bad and good overlap conditions.}
\label{tab:CI}
\begin{tabular}{ccc}
\hline
 & Bad Overlap & Good Overlap \\ \hline
Absolute Bias$ \times 1000$ &               &                \\ \hline 
Proposed                       & 60.7           & 24.6           \\ \hline
NN-covariates                  &               &                \\ \cline{1-1}
$k=1$                            & 159.6         & 57.1           \\
$k=4$                            & 432.4         & 10.3           \\
$k=16$                           & 1392.9        & 232.5          \\ \hline 
NN-PS                          &               &                \\ \cline{1-1}
$k=1$                            & 37.1          & 74.1           \\
$k=4$                            & 262.5         & 2.8            \\
$k=16$                           & 1244.6        & 52.4           \\ \hline 
IPW                            & 204.4         & 19.5           \\ \hline \hline 
Variance $\times n$                        &               &                \\ \hline
Proposed                       & 4512.6 
           & 673.3 
            \\ \hline 
NN-Covariates                  &               &                \\ \cline{1-1}
$k=1$                            & 3355.9 
           & 853.2 
            \\
$k=4$                            & 1807.6 
           & 682.2 
            \\
$k=16$                           & 1000.0 
           & 625.0 
            \\ \hline 
NN-PS                          &               &                \\ \cline{1-1}
$k=1$                            & 5245.5 
           & 1054.1 
            \\
$k=4$                            & 2767.8 
           & 807.5 
            \\
$k=16$                           & 1650.3 
           &715.6 
            \\ \hline 
IPW                            & 3298.4 
           & 818.7 
           \\ \hline
\end{tabular}
\end{table}

\section{Concluding Remarks}
\label{sec:6}

In this paper, we studied a kernel-matching estimator for treatment-effect estimation in observational studies. We established the consistency and asymptotic normality of the estimator and examined its finite-sample performance through Monte Carlo simulations across a range of settings. In these experiments, the estimator performed competitively, often on par with, and sometimes favorably relative to, covariate- and propensity-score–based matching, IPW, and DR estimators, particularly in more complex scenarios. 
We further derived the asymptotic distributions of the ATE and ATT estimators when the propensity score is estimated rather than treated as known. For the ATE, propensity score estimation yields a non-positive adjustment to the asymptotic variance; thus, matching on the estimated propensity score is at least as efficient as matching on the true score and is strictly more efficient when the adjustment is negative. For the ATT estimator, no uniform efficiency ordering can be established. We also identify two routes to semiparametric efficiency: kernel matching attains the efficiency bound under suitable sufficient-dimension-reduction conditions, while its doubly robust extension attains the bound under a more general sufficient-dimension-reduction framework.

A notable feature is the reliability of interval estimation, as evidenced by 
coverage rates that were generally reasonable and often close to the nominal 95\% level, although some undercoverage occurred in several ATT settings at smaller sample sizes. This reliability, combined with competitive bias and RMSE performance, highlights the potential of the proposed estimator for both point and interval estimation in a wide range of applications.
These results support kernel matching as a robust and flexible option for both point and interval estimation, complementing existing approaches rather than replacing them.

We also note practical considerations. Kernel matching involves computing weights for numerous pairwise comparisons, which can be computationally demanding for large datasets. Performance depends on standard tuning choices (e.g., kernel and bandwidth). Future work includes scalable implementations and principled bandwidth selection, as well as extensions to settings such as dynamic treatment regimes.

\section*{Conflict of interest}
The authors declare that there are no competing interests.

\bibliographystyle{unsrt}  
\bibliography{references}  
\section*{Appendix}
\section*{A. Proof of Main Results}

\renewcommand{\thesection}{\arabic{section}}
\setcounter{equation}{0}
\renewcommand{\theequation}{A.\arabic{equation}}
\setcounter{table}{0}
\renewcommand{\thetable}{A.\arabic{table}}

\subsection*{Preliminary:}

The influence function \citep{hampel1974influence} is crucial in analyzing the robustness of statistical estimators. Let $T$ be a real-valued functional defined on a subset of the set of all probability measures on $\mathbb{R}$, and let $F$ denote a probability measure on $\mathbb{R}$ for which $T$ is defined. 
Then, the influence function for the functional  $T(\cdot)$ is defined as
$$
\Omega(y)=\left.\frac{d}{d \epsilon} T\left[(1-\epsilon) F+\epsilon \delta_y(x)\right]\right|_{\epsilon=0},
$$
where $0<\epsilon<1$ and
$$
\delta_y(x)= \begin{cases}0 & x<y \\ 1 & x \geqslant y .\end{cases}
$$
\begin{lemma}[\cite{huber1981robust,bickel1993efficient}]
Let $X_1, \dots, X_N$ be independently and identically distributed random variables from a distribution $F$, and let $T(F)$ be a function of $F$.  Suppose $T(F)$ has a G\^{a}teaux derivative, denoted by $\Omega(x)$ and  $E(\Omega^2(X_i))<\infty$, then we have
$$
\sqrt{N}\left(T\left(F_{N}\right)-T(F)\right)=\frac{1}{\sqrt{N}} \sum_{i=1}^{N} \Omega\left(X_{i}\right)+o_{p}(1).
$$
Hence, $\sqrt{N}\left(T\left(F_{N}\right)-T(F)\right)$ is asymptotically normal with mean 0 and variance $E(\Omega^2(X_i))$.
\end{lemma}

\subsection*{Proof of Theorem 1 (i):}
We express $\hat{\tau}$ as
\begin{eqnarray*}
\hat{\tau} & = &\frac{1}{N} \sum_{i=1}^N W_i Y_i
-\frac{1}{N} \sum_{i=1}^N W_i \frac{\sum_{j=1}^N\left(1-W_j\right) Y_j K\left(\frac{p\left(X_j\right)-p\left(X_i\right)}{h}\right)}{\sum_{j=1}^N\left(1-W_j\right) K\left(\frac{p\left(X_j\right)-p\left(X_i\right)}{h}\right)} \\
& & +\frac{1}{N} \sum_{i=1}^N\left(1-W_i\right) \frac{\sum_{j=1}^N W_j Y_j K\left(\frac{p\left(X_j\right)-p\left(X_i\right)}{h}\right)}{\sum_{j=1}^N W_j K\left(\frac{p\left(X_j\right)-p\left(X_i\right)}{h}\right)}
-\frac{1}{N} \sum_{i=1}^N\left(1-W_i\right) Y_i\\
& = &T_1\left(F_N\right)-T_2\left(F_N\right)+T_3\left(F_N\right)-T_4\left(F_N\right) ,
\end{eqnarray*}
where
\begin{eqnarray*}
T_1\left(F_N\right) & = & \frac{1}{N} \sum_{i=1}^N W_i Y_i
=\int y w \mathrm{~d} F_N(y, x, w),\\
T_2\left(F_N\right)& = & \frac{1}{N} \sum_{i=1}^N W_i \frac{\sum_{j=1}^N\left(1-W_j\right) Y_j K\left(\frac{p\left(X_j\right)-p\left(X_i\right)}{h}\right)}{\sum_{j=1}^N\left(1-W_j\right) K\left(\frac{p\left(X_j\right)-p\left(X_i\right)}{h}\right)}\\
& = & \int z \frac{\int(1-w) y K\left(\frac{p(x)-p(u)}{h}\right) \mathrm{d} F_N(y, x, w)}{\int(1-w) K\left(\frac{p(x)-p(u)}{h}\right) \mathrm{d} F_N(y, x, w)} \mathrm{d} F_N(s, u, z).
\end{eqnarray*}  The terms $T_3\left(F_N\right)$ and $T_4\left(F_N\right)$ can be defined similarly.
Let $T(F) = T_1(F) - T_2(F)+T_3(F) - T_4(F),$ where
the terms $T_i(F)$ for $i = 1, 2$ are given by
\begin{eqnarray*}
T_1(F) & = & E(W Y), \\
T_2(F) & = & E\left[Z \frac{E\left[\left.(1-W) Y K\left(\frac{p(X)-p(U)}{h}\right) \right\rvert\, U\right]}{E\left[\left.(1-W) K\left(\frac{p(X)-p(U)}{h}\right) \right\rvert\, U\right]}\right].
\end{eqnarray*}    
Here, \((S, U, Z)\) has the same distribution as \((Y, X, W)\), and \((Y, X, W) \sim F_{Y,X,W}\).\\
Next, we will provide the formula for each term of $T(F)$.
$$
\begin{aligned}
& \frac{E\left[Y(1-W) K\left(\frac{p(X)-p(u)}{h}\right)\right]}{E\left[(1-W) K\left(\frac{p(X)-p(u)}{h}\right)\right]} \\
= & \bar{\mu}_0\{p(u)\}+\frac{h^2 \mu_2(K)}{2}\left[\bar{\mu}_0^{\prime \prime}\{p(u)\}+\frac{2 \bar{\mu}_0^{\prime}\{p(u)\} f_{p(X) \mid W=0}^{\prime}\{p(u)\}}{f_{p(X) \mid W=0}\{p(u)\}}\right]+o\left(h^2\right) .
\end{aligned}
$$
Then,
$$
\begin{aligned}
T_2(F)= & E\left[Z \frac{E\left[\left.Y(1-W) K\left(\frac{p(X)-p(U)}{h}\right) \right\rvert\, U\right]}{E\left[\left.(1-W) K\left(\frac{p(X)-p(U)}{h}\right) \right\rvert\, U\right]}\right] \\
= & E\left[Z \left\{\bar{\mu}_0\{p(U)\}\right.\right. \\
& \left.\left.\quad+\frac{h^2}{2}\left(\int t^2 K(t) \mathrm{d} t\right)\left[\bar{\mu}_0^{\prime \prime}\{p(U)\}+\frac{2 \bar{\mu}_0^{\prime}\{p(U)\} f_{p(X) \mid W=0}^{\prime}\{p(U)\}}{f_{p(X) \mid W=0}\{p(U)\}}\right]\right\}\right]+o\left(h^2\right) \\
= & E\left[Z \bar{\mu}_0\{p(U)\}\right]+h^2 E\left[Z b_0\{p(U)\}\right]+o\left(h^2\right) \\
= & \operatorname{Pr}(Z=1) E\{Y(0) \mid Z=1\} \\
& +h^2 \operatorname{Pr}(Z=1) E\left[b_0\{p(U)\} \mid Z=1\right]+o\left(h^2\right),
\end{aligned}
$$
where $b_0\{p(U)\}=\frac{1}{2}\left\{\int t^2 K(t) \mathrm{d} t\right\}\left[\bar{\mu}_0^{\prime \prime}\{p(U)\}+\frac{2 \bar{\mu}_0^{\prime}\{p(U)\} f_{p(X) \mid W=0}^{\prime}\{p(U)\}}{f_{p(X) \mid W=0}\{p(U)\}}\right]$.
Similarly, it follows that
$$
\begin{aligned}
T_3(F)= & E\left[( 1 - Z ) \left\{\bar{\mu}_1\{p(U)\}\right.\right. \\
& \left.\quad+\frac{h^2}{2}\left(\int t^2 K(t) \mathrm{d} t\right)\left[\bar{\mu}_1^{\prime \prime}\{p(U)\}+\frac{2 \bar{\mu}_1^{\prime}\{p(U)\} f_{p(X) \mid W=1}^{\prime}\{p(U)\}}{f_{p(X) \mid W=1}\{p(U)\}}\right]\right\}+o\left(h^2\right) \\
= & E\left[(1-Z) \bar{\mu}_1\{p(U)\}\right]+h^2 E\left[(1-Z) b_1\{p(U)\}\right]+o\left(h^2\right) \\
= & \operatorname{Pr}(Z=0) E\{Y(1) \mid Z=0\} \\
& +h^2 \operatorname{Pr}(Z=0) E\left[b_1\{p(U)\} \mid Z=0\right]+o\left(h^2\right),
\end{aligned}
$$
where $b_1\{p(U)\}=\frac{1}{2}\left(\int t^2 K(t) \mathrm{d} t\right)\left[\bar{\mu}_1^{\prime \prime}\{p(U)\}+\frac{2 \bar{\mu}_1^{\prime}\{p(U)\} f_{p(X) \mid W=1}^{\prime}\{p(U)\}}{f_{p(X) \mid W=1}\{p(U)\}}\right]$.\\
Thus,
$$
  T(F)=  T_1(F) - T_2(F) + T_3(F) - T_4(F)
  =\tau+B h^2+o(h^2),
$$
where $B=-\operatorname{Pr}(Z=1) E\left[b_0\{p(U)\} \mid Z=1\right]+\operatorname{Pr}(Z=0) E\left[b_1\{p(U)\} \mid Z=0\right]$.

To prove Theorem 1, we derive the influence function for $T(F)$ by the method presented in \cite{hampel1974influence}, \cite{hettmansperger1984statistical}, and \cite{hines2022demystifying}. 
The influence function of \(T_i(F)\), for \(i = 1, 2, 3, 4\), are given by the following expressions, respectively. After some algebraic manipulations, we obtain
\begin{eqnarray*}
\Omega_1(w, y) & = & wy - E(WY)\\
               & = & wy-E(p(X)\mu_1(X)),
\end{eqnarray*}
\begin{eqnarray*} 
\Omega_2(y, x, w) & = & \, w \frac{E[Y(1-W) K\left(\frac{X-x}{h}\right)]}{E[(1-W) K\left(\frac{X-x}{h}\right)]} - E\left[Z \frac{E[Y(1-W) K\left(\frac{X-U}{h}\right)]}{E[(1-W) K\left(\frac{X-U}{h}\right)]}\right] \\
                  & & + (1-w) y E\left[Z \frac{K\left(\frac{x-U}{h}\right)}{E[(1-W) K\left(\frac{X-U}{h}\right)]}\right] \\
                 & & - (1-w) E\left[Z \frac{E[Y(1-W) K\left(\frac{X-U}{h}\right) K\left(\frac{x-U}{h}\right)]}{\left(E[(1-W) K\left(\frac{X-U}{h}\right)]\right)^2}\right]\\
                 & = & w\mu_0(x)-E(p(X)\mu_0(X))+(1 - w)y\frac{p(x)}{1-p(x)}-(1-w)\frac{p(x)}{1-p(x)}\mu_0(x),  \\           
\Omega_3(x, y, w) & = & \, (1-w) \frac{E[WY K\left(\frac{X-x}{h}\right)]}{E[W K\left(\frac{X-x}{h}\right)]} - E\left[(1-Z) \frac{E[WY K\left(\frac{X-U}{h}\right)]}{E[W K\left(\frac{X-U}{h}\right)]}\right] \\
                 & & + wy E\left[(1-Z) \frac{K\left(\frac{x-U}{h}\right)}{E[W K\left(\frac{X-U}{h}\right)]}\right] \\
                 & & - w E\left[(1-Z) \frac{E[YW K\left(\frac{X-U}{h}\right) K\left(\frac{x-U}{h}\right)]}{\left(E[W K\left(\frac{X-U}{h}\right)]\right)^2}\right]\\
                & = &(1- w)\mu_1(x)-E((1-p(X))\mu_1(X))+wy\frac{1-p(x)}{p(x)}-w\frac{1-p(x)}{p(x)}\mu_1(x),\\
 \Omega_4(w, y) & = & (1-w) y - E[(1-W)Y]\\
                & = & (1-w) y - E[(1-p(X))\mu_0(X)].
\end{eqnarray*}
Therefore, by Lemma 7 and Markov's inequality, when $h\to 0$, we can obtain
$$
\begin{aligned}
\hat{\tau}-\tau=&T\left(F_{n}\right)-T\left(F\right)+B h^2+o(h^2)\\
=&\frac{1}{n} \sum_{i=1}^{n}\Omega\left(p(X_{i}),W_{i},Y_{i}\right)+B h^2+o(h^2) \xrightarrow{p} 0, 
\end{aligned}    
$$
where  
$\Omega\left(p(X_{i}),W_{i},Y_{i}\right)= \Omega_1\left(W_{i},Y_{i}\right)-\Omega_2\left(p(X_{i}),W_{i},Y_{i}\right)+ \Omega_3\left(p(X_{i}),W_{i},Y_{i}\right)- \Omega_4\left(W_{i},Y_{i}\right)$. 

\subsection*{Proof of Theorem 1 (ii):}

Based on the kernel matching approach for estimating the treatment effect, we express the estimator for the ATT as follows:
$$
\begin{aligned} \widehat{\tau}_{t, N} & =\frac{1}{N_1} \sum_{i=1}^N W_i\left[Y_i-\frac{\sum_{j=1}^N\left(1-W_j\right) Y_j K\left(\frac{p\left(X_j\right)-p\left(X_i\right)}{h}\right)}{\sum_{j=1}^N\left(1-W_j\right) K\left(\frac{p\left(X_j\right)-p\left(X_i\right)}{h}\right)}\right] \\ & =\frac{N}{N_1} \frac{1}{N} \sum_{i=1}^N W_i\left[Y_i-\frac{\sum_{j=1}^N\left(1-W_j\right) Y_j K\left(\frac{p\left(X_j\right)-p\left(X_i\right)}{h}\right)}{\sum_{j=1}^N\left(1-W_j\right) K\left(\frac{p\left(X_j\right)-p\left(X_i\right)}{h}\right)}\right],
\end{aligned}
$$
Let
$
B_t=-E\left[b_0\{p(U)\} \mid Z=1\right].
$
From Theorem 1(i), the population smoothing expansion satisfies
$$
T_{2, h}(F)=T_2(F)-\pi B_t h^2+o\left(h^2\right),
$$
where
$
\pi=E(W)=E\{p(X)\},
$
and
$
\tau_t=\frac{T_1(F)-T_2(F)}{\pi} .
$
The kernel matching estimator of the ATT can be expressed as
$$
\widehat{\tau}_{t, N}=\frac{T_1\left(F_N\right)-T_{2, h}\left(F_N\right)}{P_N(W)},
$$
where
$
P_N(W)=\frac{1}{N} \sum_{i=1}^N W_i=\frac{N_1}{N} .
$
Define
$
T_{N, h}=T_1\left(F_N\right)-T_{2, h}\left(F_N\right), \quad T_h=T_1(F)-T_{2, h}(F),
$
and
$
T=T_1(F)-T_2(F)=\pi \tau_t .
$
The smoothing expansion for $T_{2, h}(F)$ implies that
$$
\begin{aligned}
T_h-T & =T_2(F)-T_{2, h}(F) \\
& =\pi B_t h^2+o\left(h^2\right) .
\end{aligned}
$$
Applying the delta method to the ratio estimator gives
$$
\begin{aligned}
\frac{T_{N, h}}{P_N(W)}-\frac{T}{\pi}= & \frac{1}{\pi}\left(T_{N, h}-T_h\right)-\frac{T}{\pi^2}\left\{P_N(W)-\pi\right\} \\
& +\frac{T_h-T}{\pi}+o_p\left(N^{-1 / 2}\right)+o\left(h^2\right) .
\end{aligned}
$$
Because
$
\frac{T}{\pi^2}=\frac{\tau_t}{\pi}
$
and
$
\frac{T_h-T}{\pi}=B_t h^2+o\left(h^2\right),
$
we obtain
$$
\begin{aligned}
\widehat{\tau}_{t, N}-\tau_t= & \frac{1}{\pi}\left[\left\{T_1\left(F_N\right)-T_1(F)\right\}\right. \\
& \left.\quad-\left\{T_{2, h}\left(F_N\right)-T_{2, h}(F)\right\}\right] \\
& \quad-\frac{\tau_t}{\pi}\left\{P_N(W)-\pi\right\}+B_t h^2 \\
& \quad+o_p\left(N^{-1 / 2}\right)+o\left(h^2\right) .
\end{aligned}
$$
Equivalently, the bias-centered expansion is
$$
\begin{aligned}
\widehat{\tau}_{t, N}-\tau_t-B_t h^2=\frac{1}{\pi} & {\left[\left\{T_1\left(F_N\right)-T_1(F)\right\}\right.} \\
& \left.\quad-\left\{T_{2, h}\left(F_N\right)-T_{2, h}(F)\right\}\right] \\
& -\frac{\tau_t}{\pi}\left\{P_N(W)-\pi\right\}+o_p\left(N^{-1 / 2}\right)+o\left(h^2\right) .
\end{aligned}
$$
Finally, by Lemma 7, Markov's inequality, and the law of large numbers,
$$
P_N(W)=\frac{N_1}{N} \xrightarrow{p} \pi=E\{p(X)\}>0,
$$
Because $h \rightarrow 0$, the smoothing bias $B_t h^2$ converges to zero. Therefore,
$$
\widehat{\tau}_{t} \xrightarrow{p} \tau_t .
$$

\subsection*{Proof of Theorem 2 (i):}

According to Lemma 7 and Theorem 1, we have
\begin{equation*}
\sqrt{N}(\hat{\tau}-Bh^2-\tau)=\frac{1}{\sqrt{N}} \sum_{i=1}^{N} \Omega\left(p(X_{i}),W_{i},Y_{i}\right)+o_{p}(1).    
\end{equation*}
Since $\Omega\left(p(X_{i}),W_{i},Y_{i}\right)$ are independent and identically distributed, applying the central limit theorem, we can obtain 
\begin{equation*}
     \sqrt{N}(\hat{\tau}-Bh^2-\tau)\xrightarrow{d} \mathcal{N}(0,\sigma^2), 
\end{equation*}
Since $\Omega\left(p(X_{i}),W_{i},Y_{i}\right)$ are independent and identically distributed, applying the central limit theorem, we can obtain 
$\sigma^2=E\left(\Omega^2\left(p(X_{i}),W_{i},Y_{i}\right) \right)$, which can be expressed as 
\begin{equation}
\label{eq:KM_ATE}
\sigma^2=E\left[\frac{\bar{\sigma}_1^2(p(X))}{p(X)}+\frac{\bar{\sigma}_0^2(p(X))}{1-p(X)}+\{\tau(p(X))-\tau\}^2\right].    
\end{equation}

\subsection*{Proof of Theorem 2 (ii):}
According to Lemma 7 and Theorem 1, we can obtain
\[
\begin{aligned}
\sqrt{N}\bigl(\hat{\tau}_t-\tau_t-B_t h^2\bigr)
={}&
\frac{1}{\pi}\frac{1}{\sqrt{N}}
\sum_{i=1}^N
\Bigl[
\Omega_1(W_i,Y_i)
-\Omega_2(p(X_{i}),W_i,Y_i)
-\tau_t(W_i-\pi)
\Bigr]
+o_p(1).
\end{aligned}
\]
Applying the central limit theorem, we have
$$
\sqrt{N}(\hat{\tau}_t - B_t h^2 - \tau_t)\xrightarrow{d} N(0,\sigma_t^2),
$$
where 
\begin{align}
\sigma_t^2 &=\frac{1}{\pi^2} E\left[\Omega_1(W, Y)-\Omega_2(p(X), W, Y)-\tau_t(W-\pi)\right]^2 \\ \notag
&=\frac{1}{\pi^2} E\left[p(X) \bar{\sigma}_1^2(p(X))+\frac{p^2(X)}{1-p(X)} \bar{\sigma}_0^2(p(X))+p(X)\left(\tau(p(X))-\tau_t\right)^2\right].
\end{align}

\subsection*{Proof of Theorem 5 (i):}
Based on the observed data $\left\{\left(W_i, X_i\right)\right\}_{i=1}^N$, let the score function and the Fisher information matrix of $\beta$ be, respectively, 
$$
S(\beta)=\frac{1}{N} \sum_{i=1}^N X_i \frac{W_i-F\left(X_i^{\top} \beta\right)}{F\left(X_i^{\top} \beta\right)\left\{1-F\left(X_i^{\top} \beta\right)\right\}} f\left(X_i^{\top} \beta\right), 
$$
$$
\mathcal{I}(\beta)=E\left[\frac{f\left(X^{\top} \beta\right)^2}{F\left(X^{\top} \beta\right)\left\{1-F\left(X^{\top} \beta\right)\right\}} X X^{\top}\right],
$$
where $f(t)=\operatorname{dF}(t) / \mathrm{d} t$. Because $\hat{\beta}$ is the solution to the score equation $S(\beta)=0$, under certain regularity conditions, $\hat{\beta}-\beta^*=\mathcal{I}^{-1}_{\beta^*}S\left(\beta^*\right)+o_p\left(N^{-1 / 2}\right)$. By Taylor's series expansion, we can express
$$
\begin{aligned}
\sqrt{N}\left\{\widehat{\tau}(\widehat{\beta})-\tau-B h^2\right\}= & \frac{1}{\sqrt{N}} \sum_{i=1}^N \Omega_i  +\widetilde{T}(F) I_{\beta^*}^{-1} \frac{1}{\sqrt{N}} \sum_{i=1}^N s_i+o_p(1) .
\end{aligned}
$$
According to Lemma 7 and the central limit theorem, and $E\left(s_i s_i^{\top}\right)=I_{\beta^*}$, we obtain
$$
\begin{aligned}
\widetilde{\sigma}^2
= & \sigma^2 +\widetilde{T}(F) I_{\beta^*}^{-1}  \widetilde{T}(F)^{\top} \\
& +2 E\left(\Omega_i s_i^{\top}\right) I_{\beta^*}^{-1} \widetilde{T}(F)^{\top}.
\end{aligned}
$$
We first establish that
$E\left(\Omega_i s_i^{\top}\right)=-\widetilde{T}(F).$
To establish this result, we derive \(\widetilde{T}(F)\) and 
\(E(\Omega_i s_i^{\top})\) separately.
Let $p_\beta(x)=F(X^{\top}\beta)$, $p(x)=p_{\beta^*}$.
 $
\dot{p}_\beta(x)=\frac{\partial p_\beta(x)}{\partial \beta}=f\left(x^{\top} \beta\right) x .
$
Define $R=p(X)$,  $R_{U}=p(U)$, 
$
m_{0 h}(U ; \beta)=\frac{A_{0 h}(U ; \beta)}{B_{0 h}(U ; \beta)},
$
where 
$A_{0 h}(U ; \beta)  =E\left[(1-W) Y K\left(\frac{p_{\beta}(X)-p_{\beta}(U)}{h}\right)\right]$, 
$B_{0 h}(U ; \beta)  =E\left[(1-W) K\left(\frac{p_{\beta}(X)-p_{\beta}(U)}{h}\right)\right]$ .
Therefore,
\begin{equation}
\label{eq:TF}
\widetilde{T}_{1 h}(F)=-E\left[\left.Z \frac{\partial m_{0 h}(U ; \beta)}{\partial \beta^{\top}}\right|_{\beta=\beta^{*}}\right] .
\end{equation}
For any $a\in\mathbb{R}^q$, let $D_a$ denote
the total directional derivative at $\beta^*$ and define, for
$V\in\{X,U\}$,
\[
R_V=p_{\beta^*}(V),
\qquad
Q(V)=\dot p(V)^\top a,
\qquad
\dot p(V)=f(V^\top\beta^*)V.
\]
By Eq. \eqref{eq:TF}, we have
\begin{equation}
\widetilde T_{1h}(F)a
=
-E\!\left[Z D_a m_{0h}(U)\right],
\end{equation}
where
\begin{align}
D_a m_{0h}(U)
&=
\frac{
 E\!\left[
 (1-R)\{\mu_0(X)-m_{0h}(U)\}
 K'\!\left(\frac{R-R_U}{h}\right)
 \frac{Q(X)-Q(U)}{h}
 \right]
}{
 E\!\left[
 (1-R)K\!\left(\frac{R-R_U}{h}\right)
 \right]
}.
\label{eq:directional-derivative-m0h}
\end{align}
Let \(g\) denote the density of \(R\), and define
\[
b(r)=(1-r)g(r),\qquad
m_0(r)=E\{\mu_0(X)\mid R=r\},
\]
\[
\bar q(r)=E\{Q(X)\mid R=r\},\qquad
c_0(r)=\operatorname{Cov}\{\mu_0(X),Q(X)\mid R=r\}.
\]
For \(r=R_U\) and \(q=Q(U)\),
\[
E\!\left[
 \{\mu_0(X)-m_{0h}(r)\}\{Q(X)-q\}\mid R=v
\right]
=
c_0(v)+\{m_0(v)-m_{0h}(r)\}\{\bar q(v)-q\}.
\]
Consequently, Eq. \eqref{eq:directional-derivative-m0h} can be written as
\[
D_a m_{0h}(U)=\frac{N_h(r,q)}{B_h(r)},
\]
where
\begin{align*}
N_h(r,q)
&=
\int b(v)
\left[
c_0(v)+\{m_0(v)-m_{0h}(r)\}\{\bar q(v)-q\}
\right]
K'\!\left(\frac{v-r}{h}\right)\frac{dv}{h},\\
B_h(r)
&=
\int b(v)K\!\left(\frac{v-r}{h}\right)dv.
\end{align*}
Under the stated smoothness, overlap, kernel, and boundary conditions,
the change of variables \(v=r+ht\) gives, uniformly for interior \(r\),
\[
B_h(r)=hb(r)+O(h^3),
\qquad
m_{0h}(r)=m_0(r)+O(h^2).
\]
Assume  that $K$ is a symmetric, continuously differentiable second-order kernel satisfying
$
\int K(t) d t=1, \quad \int t K(t) d t=0, \quad \int K^{\prime}(t) d t=0, \quad \int t K^{\prime}(t) d t=-1
$,  the numerator satisfies
\[
N_h(r,q)
=
-h\left[
\{b(r)c_0(r)\}'
+b(r)m_0'(r)\{\bar q(r)-q\}
\right]
+o(h).
\]
It follows that
\begin{equation*}
D_a m_{0h}(U)
=
-\frac{\{b(R_U)c_0(R_U)\}'}{b(R_U)}
-m_0'(R_U)\{\bar q(R_U)-Q(U)\}
+o(1).
\end{equation*}
Therefore,
\begin{align*}
\widetilde T_{1h}(F)a
&=
E\!\left[
R_U\frac{\{b(R_U)c_0(R_U)\}'}{b(R_U)}
\right]\\
&\quad+
E\!\left[
R_Um_0'(R_U)\{\bar q(R_U)-Q(U)\}
\right]
+o(1).
\end{align*}
The second expectation is equal to zero because
\(E\{Q(U)\mid R_U\}=\bar q(R_U)\). Hence, letting
\(h\to0\) and integrating by parts,
\begin{align}
\widetilde T_1(F)a
&=
\int \frac{r}{1-r}\{b(r)c_0(r)\}'\,dr
\notag\\
&=
-\int\frac{b(r)c_0(r)}{(1-r)^2}\,dr
=
-E\!\left\{\frac{c_0(R)}{1-R}\right\},
\label{eq:T1-directional-limit}
\end{align}
where the boundary term is zero by assumption.
Since
$c_0(R)
=
\operatorname{Cov}\{\mu_0(X),\dot p(X)^\top\mid R\}a$
and \(a\) is arbitrary, Eq. \eqref{eq:T1-directional-limit} implies
\begin{equation*}
\widetilde T_1(F)
=
-E\!\left[
\frac{
\operatorname{Cov}\{\mu_0(X),\dot p(X)^\top\mid p(X)\}
}{
1-p(X)
}
\right].
\end{equation*}
Next, we calculate $ E(\Omega_0s^\top)$. Let
\[
\Omega_0
=
-\frac{1-W}{1-R}\{Y-\bar\mu_0(R)\},
\qquad
s
=
\frac{W-R}{R(1-R)}\dot p(X),
\]
where \(\bar\mu_0(R)=m_0(R)\). For every \(a\in\mathbb R^q\),
\[
E(\Omega_0s^\top a\mid X)
=
\frac{Q(X)\{\mu_0(X)-\bar\mu_0(R)\}}{1-R}.
\]
Conditioning once more on \(R\) gives
\[
E(\Omega_0s^\top a)
=
E\!\left\{\frac{c_0(R)}{1-R}\right\}
=
-\widetilde T_1(F)a.
\]
Since this holds for every \(a\),
\begin{equation*}
E(\Omega_0s^\top)=-\widetilde T_1(F).
\label{eq:score-T1-identity}
\end{equation*}
Similarly, we obtain
\begin{equation*}
\widetilde{T}_{2}(F)=-E\left[\frac{\operatorname{Cov}\left\{\mu_{1}(X), \dot{p}(X)^{\top} \mid p(X)\right\}}{p(X)}\right],
\end{equation*}
and
\begin{equation*}
E\left(\Omega_{1 i} s_{i}^{\top}\right)=-\widetilde{T}_{2}(F) .
\end{equation*}
Therefore,
\begin{equation*}
E\left(\Omega_{i} s_{i}^{\top}\right)=-\left\{\widetilde{T}_{1}(F)+\widetilde{T}_{2}(F)\right\}=-\widetilde{T}(F) . 
\end{equation*}
Consequently, the asymptotic variance of the ATE estimator using the estimated propensity score can be written as
$$
\tilde{\sigma}^2=\sigma^2-\tilde{T}(F) I_{\beta^*}^{-1} \widetilde{T}(F)^{\top}
$$

\subsection*{Proof of Theorem 5 (ii):}
From the preceding expansion,
$$
\begin{aligned}
& \sqrt{N}\left\{\widehat{\tau}_{t, N}(\widehat{\beta})-\tau_t-B_t h^2\right\} \\
& =\frac{1}{\pi \sqrt{N}} \sum_{i=1}^N\left[\Omega_1\left(W_i, Y_i\right)-\Omega_2\left(X_i, W_i, Y_i\right)-\tau_t\left(W_i-\pi\right)\right. \\
& \quad+\widetilde{T}_{1}(F)
 \times I^{-1}_{\beta^*}  \left.  \frac{W_i-F\left(X_i^{\top} \beta^*\right)}{F\left(X_i^{\top} \beta^*\right)\left[1-F\left(X_i^{\top} \beta^*\right)\right]} f\left(X_i^{\top} \beta^*\right) X_i\right]+o_p(1) .
\end{aligned}
$$
The contribution corresponding to the estimator based on the true propensity
score can be written as
\[
  \Omega_{\mathrm{ATT},i}
  =
  \frac{1}{\pi}\left[
    W_i\{Y_i-\overline\mu_0(p_i)-\tau_{\mathrm{ATT}}\}
    -(1-W_i)\frac{p_i}{1-p_i}
      \{Y_i-\overline\mu_0(p_i)\}
  \right].
\]
Define
\[
\begin{aligned}
a_t
&:=
\left.
\nabla_\beta\tau_t(\beta)
\right|_{\beta=\beta^*}
\\
&=
\frac{1}{\pi}
E\left[
\dot p(X)
\{\mu_1(X)-\mu_0(X)-\tau_t\}
\right],
\end{aligned}
\]
Define the score--influence covariance vector as
$c_t
:=
E\left(s_i\Omega_{\mathrm{ATT},i}\right).$
Using the expression for $\Omega_{\mathrm{ATT},i}$ and iterated
expectations, this vector can first be written as
\[
\begin{aligned}
c_t
=
\frac{1}{\pi}E\Bigg[
&\operatorname{Cov}
\{\dot p(X),\mu_1(X)\mid p(X)\}
\\
&+
\frac{p(X)}{1-p(X)}
\operatorname{Cov}
\{\dot p(X),\mu_0(X)\mid p(X)\}
\\
&+
E\{\dot p(X)\mid p(X)\}
\{\overline\mu_1(p(X))
-\overline\mu_0(p(X))
-\tau_t\}
\Bigg].
\end{aligned}
\]
So, we obtain $\left(a_t-c_t\right)^{\top}=\frac{1}{\pi} \tilde{T}_1(F)$.
Combining this identity with the maximum-likelihood expansion for
$\widehat\beta$ gives the influence function under the estimated propensity
score:
\[
\widetilde\Omega_{\mathrm{ATT},i}
=
\Omega_{\mathrm{ATT},i}
+
(a_t-c_t)^\top
I_{\beta^*}^{-1}s_i.
\]
Therefore,
\[
\sqrt N
\left\{
\widehat\tau_{t,N}(\widehat\beta)
-\tau_t-B_t h^2
\right\}
=
\frac{1}{\sqrt N}
\sum_{i=1}^N
\widetilde\Omega_{\mathrm{ATT},i}
+o_p(1).
\]
Let
$\sigma_{t}^2
=
E\left(\Omega_{\mathrm{ATT},i}^2\right).$
Using
$E\left(s_i\Omega_{\mathrm{ATT},i}\right)=c_t$,
$E(s_i s_i^\top)=I_{\beta^*}$,
the asymptotic variance under the estimated propensity score is
\[
\begin{aligned}
\widetilde\sigma_t^2
&=
E\left(
\widetilde\Omega_{\mathrm{ATT},i}^2
\right)
\\
&=
\sigma_t^2
-c_t^\top I_{\beta^*}^{-1}c_t
+a_t^\top I_{\beta^*}^{-1}a_t.
\end{aligned}
\]

\section*{B. Simulation Results for Kernel Matching Estimators Based on the True and Estimated Propensity Scores}

\renewcommand{\thesection}{\arabic{section}}
\setcounter{equation}{0}
\renewcommand{\theequation}{B.\arabic{equation}}
\setcounter{table}{0}
\renewcommand{\thetable}{B.\arabic{table}}

\begin{table}[H]
\caption{Performance of kernel matching estimators for the ATE and ATT based on the true and estimated propensity scores under Settings I and II}
\label{tab:kernel_matching_simulation}
\centering

\begin{threeparttable}

\begin{tabular}{ccccccc}
\hline
\multicolumn{2}{c}{}           & \multicolumn{2}{c}{ATE}     &  & \multicolumn{2}{c}{ATT}     \\ \cline{3-4} \cline{6-7} 
             &                 & \multicolumn{2}{c}{Setting} &  & \multicolumn{2}{c}{Setting} \\ \cline{3-4} \cline{6-7} 
             &                 & I            & II           &  & I            & II           \\ \hline
True PS      & Bias            & -0.002       & -0.001       &  & -0.003       & -0.001       \\
             & SD              & 0.062        & 0.112        &  & 0.072        & 0.122        \\
             & SE              & 0.062        & 0.112        &  & 0.068        & 0.122        \\
             & 95\% Coverage  & 0.949        & 0.958        &  & 0.949        & 0.966        \\
Estimated PS & Bias            & -0.001       & -0.002       &  & -0.002       & -0.001       \\
             & SD              & 0.047        & 0.112        &  & 0.056        & 0.153        \\
             & SE              & 0.047        & 0.112        &  & 0.054        & 0.158        \\
             & 95\% Coverage & 0.949        & 0.956        &  & 0.945        & 0.963        \\ \hline
\end{tabular}

\begin{tablenotes}
\footnotesize
\item[] $N=3000$, bandwidth $h=0.01$, and the number of Monte Carlo replications is 1,000.
\end{tablenotes}
\end{threeparttable}
\end{table}
To complement the theoretical variance comparison, we conduct Monte Carlo experiments under Settings I and II for the kernel matching estimators based on the true and estimated propensity scores. We report the bias, SD, SE, and the empirical coverage probability of the nominal 95\% confidence
intervals, where the SE is computed using the proposed consistent variance estimator.

Table~\ref{tab:kernel_matching_simulation} provides numerical support for the theoretical conclusions. For the ATE, estimating the propensity score strictly improves precision in Setting I and leaves the empirical precision essentially unchanged in Setting II. In Setting I, the  SD decreases from 0.062 under the true propensity score to 0.047 under the estimated propensity score.  These results are consistent with the theoretical finding that estimating a correctly specified
propensity score weakly reduces the asymptotic variance of the ATE estimator. In Setting II, since $\widetilde{T}(F)=0$, estimating the propensity score has no first-order effect on the asymptotic variance. Consequently, the SD remains essentially unchanged relative to that obtained using the true propensity score.

The ATT results illustrate why no analogous universal ordering holds. In Setting I, estimating the propensity score reduces the empirical SD from 0.072 to 0.056. By contrast, in Setting II, the empirical
SD increases from 0.122 to 0.153. Thus, depending on
the data-generating process, estimating the propensity score can either decrease or increase the asymptotic variance of the ATT estimator, as predicted by the theoretical variance decomposition.

Across all cases, the biases are negligible. Moreover, the estimated SEs are close to the corresponding empirical SD, and the
coverage probabilities remain close to the nominal 95\% levels. These findings support both the validity of the proposed variance estimators and the theoretical distinction between the ATE and ATT: estimation of the propensity score weakly improves ATE efficiency, whereas its effect on ATT efficiency is design-dependent.

\section*{C. Technical Details and Simulation Results for the Comparison with Semiparametric Efficiency Bounds}

\setcounter{equation}{0}
\renewcommand{\theequation}{C.\arabic{equation}}
\setcounter{table}{0}
\renewcommand{\thetable}{C.\arabic{table}}

In this appendix, we provide the technical details and simulation evidence supporting the efficiency results summarized in the main text. We first present the semiparametric efficiency bounds for the ATE and ATT under the unconfoundedness model. We then establish two results. First, we show that propensity-score kernel matching attains the corresponding efficiency bounds when the propensity score provides sufficient dimension reduction for the conditional outcome moments. Second, we show that the proposed doubly robust kernel-matching estimators attain these bounds under a more general sufficient dimension reduction structure. We also provide consistent variance estimators, asymptotic confidence intervals, and simulation results illustrating the finite-sample implications of the theoretical findings.

Under the unconfoundedness and overlap assumptions, the semiparametric
efficiency bound for the ATE estimator is \citep{hahn1998role}
\begin{equation}
\label{eq:ATEBOUND}
V_{\mathrm{EIF}}^{\mathrm{ATE}}
=
E\left[
\frac{\sigma_1^2(X)}{p(X)}
+
\frac{\sigma_0^2(X)}{1-p(X)}
+
\{\tau(X)-\tau\}^2
\right].
\end{equation}
The corresponding efficiency bound for the ATT estimator is
\begin{equation}
\label{eq:ATTBOUND}
V_{\mathrm{EIF}}^{\mathrm{ATT}}
=
E\left[
\frac{p(X)\sigma_1^2(X)}{\pi^2}
+
\frac{p^2(X)\sigma_0^2(X)}
     {\pi^2\{1-p(X)\}}
+
\frac{p(X)\{\tau(X)-\tau_t\}^2}{\pi^2}
\right].
\end{equation}

As emphasized by \cite{hahn1998role}, the propensity score contributes
to efficient estimation primarily by providing dimension reduction,
rather than by supplying additional identifying information. We next
describe two settings in which the proposed kernel-based estimators
attain the bounds in Eqs. \eqref{eq:ATEBOUND} and \eqref{eq:ATTBOUND}.

\subsection*{C.1. Efficiency under propensity-score dimension reduction}

For \(u\in(0,1)\), define
\[
\bar{\mu}_w(u)
=
E\{Y(w)\mid p(X)=u\},
\qquad
\bar{\sigma}_w^2(u)
=
\operatorname{Var}\{Y(w)\mid p(X)=u\},
\quad w=0,1.
\]
Consider the following sufficient dimension reduction condition:
\begin{equation}
\label{eq:dr1}
\bar{\mu}_w\{p(x)\}=\mu_w(x)
\quad\text{and}\quad
\bar{\sigma}_w^2\{p(x)\}=\sigma_w^2(x),
\qquad w=0,1,
\end{equation}
for almost every \(x\). Condition \eqref{eq:dr1} requires the propensity
score to retain the information in \(X\) relevant to the first two
conditional moments of the potential outcomes. In particular, it is
satisfied when both \(\mu_w(x)\) and \(\sigma_w^2(x)\) are functions of
\(p(x)\). It is also satisfied when the propensity score and these
conditional moments depend on a common scalar index and the propensity
score is one-to-one in that index.

\begin{proposition}
\label{prop:ps-efficiency}
Suppose that the unconfoundedness, overlap, smoothness, and bandwidth
conditions imposed in the main text hold. If condition \eqref{eq:dr1}
is satisfied, then the adjusted propensity-score kernel-matching
estimators of the ATE and ATT attain the corresponding semiparametric
efficiency bounds in Eqs. \eqref{eq:ATEBOUND} and \eqref{eq:ATTBOUND}.
\end{proposition}

\begin{proof}
The asymptotic expansion of the adjusted kernel-matching ATE estimator
derived in the main text reduces, under condition \eqref{eq:dr1}, to
\[
\sqrt{N}(\widehat{\tau}-\tau)
=
\frac{1}{\sqrt{N}}
\sum_{i=1}^N
\phi_{\mathrm{ATE}}(Z_i)
+o_p(1),
\]
where \(Z_i=(Y_i,W_i,X_i)\) and
\[
\phi_{\mathrm{ATE}}(Z_i)
=
\frac{W_i\{Y_i-\mu_1(X_i)\}}{p(X_i)}
-
\frac{(1-W_i)\{Y_i-\mu_0(X_i)\}}{1-p(X_i)}
+
\tau(X_i)-\tau.
\]
This is the efficient influence function for the ATE under the
unconfoundedness model. Its variance is
\[
\operatorname{Var}\{\phi_{\mathrm{ATE}}(Z_i)\}
=
V_{\mathrm{EIF}}^{\mathrm{ATE}}.
\]

Similarly, the asymptotic expansion of the adjusted kernel-matching ATT
estimator becomes
\[
\sqrt{N}(\widehat{\tau}^{\,t}-\tau_t)
=
\frac{1}{\sqrt{N}}
\sum_{i=1}^N
\phi_{\mathrm{ATT}}(Z_i)
+o_p(1),
\]
where
\[
\begin{aligned}
\phi_{\mathrm{ATT}}(Z_i)
=
\frac{1}{\pi}\Bigg[
&
W_i\{Y_i-\mu_1(X_i)\}
-
\frac{(1-W_i)p(X_i)}
     {1-p(X_i)}
\{Y_i-\mu_0(X_i)\}
+
W_i\{\tau(X_i)-\tau_t\}
\Bigg].
\end{aligned}
\]
This is the efficient influence function for the ATT, and
\[
\operatorname{Var}\{\phi_{\mathrm{ATT}}(Z_i)\}
=
V_{\mathrm{EIF}}^{\mathrm{ATT}}.
\]
The result follows from the central limit theorem.
\end{proof}

This result is closely related to \cite{hf}, who show that matching
estimators can attain the semiparametric efficiency bound when the
matching variables provide a valid low-dimensional sufficient
representation and the effective number of matches diverges at an
appropriate rate.

\subsection*{C.2. Doubly robust kernel matching under sufficient dimension reduction}

We next consider an augmented kernel-matching estimator, equivalently
a doubly robust estimator whose nuisance functions are estimated using
kernel smoothing. Following \cite{wangjun}, suppose that the treatment
assignment mechanism and conditional outcome regression functions
admit a common sufficient dimension reduction structure.

Let
\[
R=B^{*\top}X\in\mathbb{R}^r,
\]
where \(B^*\) is a \(k\times r\) matrix and \(r\ll k\).

\begin{assumption}
\label{ass:sdr}
Suppose that the following conditions hold:
\begin{enumerate}
\item[(i)]
The potential outcomes are conditionally independent of the treatment
assignment given \(R\):
\[
\{Y(1),Y(0)\}\perp W\mid R.
\]

\item[(ii)]
The propensity score and conditional outcome regression functions
depend on \(X\) only through \(R\):
\[
p(X)=p_R(R),
\qquad
\mu_w(X)=\mu_{w,R}(R),
\quad w=0,1.
\]

\end{enumerate}
\end{assumption}

The central subspace spanned by \(B^*\) is estimated using sliced
inverse regression (SIR) \citep{Li1991}. Let \(\widehat{B}\) denote
the resulting estimator and define
\[
\widehat{R}_i=\widehat{B}^{\top}X_i,
\qquad
\widehat{R}(x)=\widehat{B}^{\top}x.
\]
The unknown link functions are estimated nonparametrically using kernel
regression. In particular,
\begin{equation}
\label{eq:p-index-est}
\widehat{p}(x)
=
\frac{
\displaystyle
\sum_{i=1}^N
K_{h_p}\{\widehat{R}_i-\widehat{R}(x)\}W_i
}{
\displaystyle
\sum_{i=1}^N
K_{h_p}\{\widehat{R}_i-\widehat{R}(x)\}
},
\end{equation}
where \(K_{h_p}(u)=h_p^{-r}K(u/h_p)\). Similarly,
\begin{equation}
\label{eq:mu-index-est}
\widehat{\mu}_w(x)
=
\frac{
\displaystyle
\sum_{i=1}^N
K_{h_w}\{\widehat{R}_i-\widehat{R}(x)\}
\mathbf{1}(W_i=w)Y_i
}{
\displaystyle
\sum_{i=1}^N
K_{h_w}\{\widehat{R}_i-\widehat{R}(x)\}
\mathbf{1}(W_i=w)
},
\qquad w=0,1.
\end{equation}

The doubly robust kernel-matching estimator of the ATE is
\begin{equation}
\label{eq:tauK_ATE}
\begin{aligned}
\widehat{\tau}_{K,\mathrm{DR}}
=
\frac{1}{N}\sum_{i=1}^N
\Bigg[
&
\frac{W_i\{Y_i-\widehat{\mu}_1(X_i)\}}
     {\widehat{p}(X_i)}
-
\frac{(1-W_i)\{Y_i-\widehat{\mu}_0(X_i)\}}
     {1-\widehat{p}(X_i)}
+
\widehat{\mu}_1(X_i)-\widehat{\mu}_0(X_i)
\Bigg].
\end{aligned}
\end{equation}
The corresponding estimator of the ATT is
\begin{equation}
\label{eq:tauK_ATT}
\begin{aligned}
\widehat{\tau}_{K,\mathrm{DR}}^{\,t}
=
\frac{1}{N\widehat{\pi}}
\sum_{i=1}^N
\Bigg[
&
W_i\{Y_i-\widehat{\mu}_1(X_i)\}
-
\frac{(1-W_i)\widehat{p}(X_i)}
     {1-\widehat{p}(X_i)}
\{Y_i-\widehat{\mu}_0(X_i)\}
\\
&+
W_i\{\widehat{\mu}_1(X_i)-\widehat{\mu}_0(X_i)\}
\Bigg],
\end{aligned}
\end{equation}
where
\[
\widehat{\pi}=\frac{1}{N}\sum_{i=1}^N W_i.
\]

\begin{theorem}
\label{th:ATEBOUND}
Suppose that Assumptions \ref{assum1}--\ref{ass:sdr} hold, and
that the SIR and kernel estimators satisfy the regularity and rate
conditions stated in the main text. Then, 
\[
\sqrt{N}
\left(
\widehat{\tau}_{K,\mathrm{DR}}-\tau
\right)
=
\frac{1}{\sqrt{N}}
\sum_{i=1}^N
\phi_{\mathrm{ATE}}(Z_i)
+o_p(1),
\]
and hence
\[
\sqrt{N}
\left(
\widehat{\tau}_{K,\mathrm{DR}}-\tau
\right)
\xrightarrow{d}
N(0,\sigma_{\mathrm{ATE}}^2),
\]
where
\begin{equation}
\label{eq:varATE}
\begin{aligned}
\sigma_{\mathrm{ATE}}^2
&=
E\left[
\frac{\sigma_1^2(X)}{p(X)}
+
\frac{\sigma_0^2(X)}{1-p(X)}
+
\{\tau(X)-\tau\}^2
\right]  =
V_{\mathrm{EIF}}^{\mathrm{ATE}}.
\end{aligned}
\end{equation}
Therefore, \(\widehat{\tau}_{K,\mathrm{DR}}\) attains the semiparametric
efficiency bound for the ATE.
\end{theorem}

\begin{proof}
Let \(\eta=(p,\mu_0,\mu_1)\) and define
\[
\begin{aligned}
\psi_{\mathrm{ATE}}(Z;\eta)
=
&
\frac{W\{Y-\mu_1(X)\}}{p(X)}
-
\frac{(1-W)\{Y-\mu_0(X)\}}{1-p(X)}
\\
&+
\mu_1(X)-\mu_0(X).
\end{aligned}
\]
Then,
$
E\{\psi_{\mathrm{ATE}}(Z;\eta_0)\}=\tau$,
where \(\eta_0\) denotes the true nuisance functions. Let \(P_N\)
denote the empirical average and \(P\) the population expectation. An
expansion around \(\eta_0\) gives
\[
\begin{aligned}
\widehat{\tau}_{K,\mathrm{DR}}-\tau
=
&
(P_N-P)\psi_{\mathrm{ATE}}(Z;\eta_0)
\\
&+
P\left[
\psi_{\mathrm{ATE}}(Z;\widehat{\eta})
-
\psi_{\mathrm{ATE}}(Z;\eta_0)
\right]
\\
&+
(P_N-P)\left[
\psi_{\mathrm{ATE}}(Z;\widehat{\eta})
-
\psi_{\mathrm{ATE}}(Z;\eta_0)
\right].
\end{aligned}
\]
The estimating function is Neyman-orthogonal with respect to the
propensity score and outcome regression functions. Under the stated
smoothness, bandwidth, and convergence-rate conditions, the last two
terms are \(o_p(N^{-1/2})\). Therefore,
\[
\sqrt{N}
\left(
\widehat{\tau}_{K,\mathrm{DR}}-\tau
\right)
=
\frac{1}{\sqrt{N}}
\sum_{i=1}^N
\left[
\psi_{\mathrm{ATE}}(Z_i;\eta_0)-\tau
\right]
+o_p(1).
\]
The term in brackets is \(\phi_{\mathrm{ATE}}(Z_i)\). Its variance is
given by Eq. \eqref{eq:varATE}, and the result follows from the central
limit theorem.
\end{proof}

\begin{theorem}
\label{th:ATTBOUND}
Under the conditions of Theorem \ref{th:ATEBOUND},
\[
\sqrt{N}
\left(
\widehat{\tau}_{K,\mathrm{DR}}^{\,t}-\tau_t
\right)
=
\frac{1}{\sqrt{N}}
\sum_{i=1}^N
\phi_{\mathrm{ATT}}(Z_i)
+o_p(1),
\]
and hence
\[
\sqrt{N}
\left(
\widehat{\tau}_{K,\mathrm{DR}}^{\,t}-\tau_t
\right)
\xrightarrow{d}
N(0,\sigma_{\mathrm{ATT}}^2),
\]
where
\begin{equation}
\label{eq:varATT}
\begin{aligned}
\sigma_{\mathrm{ATT}}^2
&=
E\left[
\frac{p(X)\sigma_1^2(X)}{\pi^2}
+
\frac{p^2(X)\sigma_0^2(X)}
     {\pi^2\{1-p(X)\}}
+
\frac{p(X)\{\tau(X)-\tau_t\}^2}{\pi^2}
\right] \\
&=
V_{\mathrm{EIF}}^{\mathrm{ATT}}.
\end{aligned}
\end{equation}
Therefore, \(\widehat{\tau}_{K,\mathrm{DR}}^{\,t}\) attains the
semiparametric efficiency bound for the ATT.
\end{theorem}

\begin{proof}
Define
\[
\begin{aligned}
m_{\mathrm{ATT}}(Z;\eta)
=
&
W\{Y-\mu_1(X)\}
-
\frac{(1-W)p(X)}{1-p(X)}
\{Y-\mu_0(X)\}
\\
&+
W\{\mu_1(X)-\mu_0(X)\}.
\end{aligned}
\]
At the true nuisance functions,
\[
E\{m_{\mathrm{ATT}}(Z;\eta_0)\}
=
\pi\tau_t.
\]
The estimator in Eq. \eqref{eq:tauK_ATT} is the ratio of an estimator of
\(\pi\tau_t\) to \(\widehat{\pi}\). By the same orthogonality argument
used for the ATE estimator,
\[
P_Nm_{\mathrm{ATT}}(Z;\widehat{\eta})
=
P_Nm_{\mathrm{ATT}}(Z;\eta_0)
+
o_p(N^{-1/2}).
\]
A first-order expansion of the ratio around
\((\pi\tau_t,\pi)\) gives
\[
\begin{aligned}
\sqrt{N}
\left(
\widehat{\tau}_{K,\mathrm{DR}}^{\,t}-\tau_t
\right)
=
\frac{1}{\sqrt{N}}
\sum_{i=1}^N
\frac{
m_{\mathrm{ATT}}(Z_i;\eta_0)-\tau_tW_i
}{\pi}
+
o_p(1).
\end{aligned}
\]
The summand is precisely \(\phi_{\mathrm{ATT}}(Z_i)\). Its variance is
given by Eq. \eqref{eq:varATT}, and the central limit theorem completes the
proof.
\end{proof}

\begin{remark}
The term ``doubly robust'' refers to the consistency of the estimators
when either the propensity score or the relevant outcome regression
functions are consistently estimated, subject to the usual regularity
conditions. Attainment of the semiparametric efficiency bound requires
both sets of nuisance functions to be consistently estimated at
sufficiently fast rates.
\end{remark}

If matching and kernel smoothing are performed using the reduced
covariate \(R\), the dimensionality burden is substantially lower than
when the full covariate vector is used. Under Assumption
\ref{ass:sdr}, the leading terms in the asymptotic expansions coincide
with the efficient influence functions. The result is related to
\cite{hfecon} and \cite{LIN2025106080}, who study diverging-\(M\)
matching based on the original covariate vector. The present approach
instead obtains efficiency by matching and smoothing over a
lower-dimensional sufficient representation.

\subsection*{C.3. Variance estimation and confidence intervals}

Consistent estimators of the asymptotic variances can be constructed
using the estimated efficient influence functions. For the ATE, define
\[
\begin{aligned}
\widehat{\phi}_{\mathrm{ATE},i}
=
&
\frac{W_i\{Y_i-\widehat{\mu}_1(X_i)\}}
     {\widehat{p}(X_i)}
-
\frac{(1-W_i)\{Y_i-\widehat{\mu}_0(X_i)\}}
     {1-\widehat{p}(X_i)}
\\
&+
\widehat{\mu}_1(X_i)-\widehat{\mu}_0(X_i)
-
\widehat{\tau}_{K,\mathrm{DR}}.
\end{aligned}
\]
For the ATT, define
\[
\begin{aligned}
\widehat{\phi}_{\mathrm{ATT},i}
=
\frac{1}{\widehat{\pi}}
\Bigg[
&
W_i\{Y_i-\widehat{\mu}_1(X_i)\}
-
\frac{(1-W_i)\widehat{p}(X_i)}
     {1-\widehat{p}(X_i)}
\{Y_i-\widehat{\mu}_0(X_i)\}
\\
&+
W_i
\left\{
\widehat{\mu}_1(X_i)-\widehat{\mu}_0(X_i)
-\widehat{\tau}_{K,\mathrm{DR}}^{\,t}
\right\}
\Bigg].
\end{aligned}
\]
Let
$
\overline{\widehat{\phi}}_{\mathrm{ATE}}
=
\frac{1}{N}\sum_{i=1}^N
\widehat{\phi}_{\mathrm{ATE},i}
\mbox{ and } \overline{\widehat{\phi}}_{\mathrm{ATT}}
=
\frac{1}{N}\sum_{i=1}^N
\widehat{\phi}_{\mathrm{ATT},i}.$
The variance estimators are
\[
\widehat{\sigma}_{\mathrm{ATE}}^2
=
\frac{1}{N}
\sum_{i=1}^N
\left(
\widehat{\phi}_{\mathrm{ATE},i}
-
\overline{\widehat{\phi}}_{\mathrm{ATE}}
\right)^2
\]
and
\[
\widehat{\sigma}_{\mathrm{ATT}}^2
=
\frac{1}{N}
\sum_{i=1}^N
\left(
\widehat{\phi}_{\mathrm{ATT},i}
-
\overline{\widehat{\phi}}_{\mathrm{ATT}}
\right)^2.
\]

An asymptotic \(100(1-\alpha)\%\) confidence interval for the ATE is
\begin{equation}
\label{eq:CI}
\widehat{\tau}_{K,\mathrm{DR}}
\pm
z_{1-\alpha/2}
\frac{\widehat{\sigma}_{\mathrm{ATE}}}{\sqrt{N}}.
\end{equation}
The corresponding confidence interval for the ATT is
\[
\widehat{\tau}_{K,\mathrm{DR}}^{\,t}
\pm
z_{1-\alpha/2}
\frac{\widehat{\sigma}_{\mathrm{ATT}}}{\sqrt{N}}.
\]

\subsection*{C.4. Simulation evidence}

We conduct simulation studies to examine whether the finite-sample
variability of the proposed estimators approaches the corresponding
semiparametric efficiency bounds and whether the influence-function
confidence intervals provide reliable coverage. The two settings are
adapted from \cite{abadie2016matching} and \cite{hf}; both have a
true ATE of \(5\). 
In each setting, we impose the sufficient dimension reduction structure and perform kernel matching based on the reduced index $X^\top\beta$. Under condition (E.3), this index is sufficient for both treatment assignment and outcome regression. Consequently, kernel matching based on $X^\top\beta$ may attain the semiparametric efficiency bound.

For the ATE, the square roots of the semiparametric efficiency bounds,
denoted by \(\sigma^{\mathrm{eff}}\), are \(2.473\) and \(6.139\) under
Settings I and II, respectively. For the ATT, the corresponding values
are \(2.816\) and \(8.449\). We consider sample sizes
\(N=200,500,\) and \(1000\), with \(1{,}000\) Monte Carlo replications
for each design.

We report the mean absolute error (MAE), root mean squared error
(RMSE), average width of the \(95\%\) confidence interval (AW),
empirical coverage probabilities of the \(90\%\) and \(95\%\)
confidence intervals, and normalized standard deviation (NSD), defined
as
\[
\mathrm{NSD}
=
\sqrt{N}\,
\operatorname{SD}_{\mathrm{MC}}(\widehat{\tau}).
\]
The NSD can therefore be compared directly with
\(\sigma^{\mathrm{eff}}\).

\begin{table}[H]
\centering
\small
\caption{Simulation results for the propensity-score kernel-matching ATE estimator under Setting I}
\begin{tabular}{ccccccc}
\hline
Sample size & MAE & RMSE & AW & 90\% Coverage & 95\% Coverage & NSD \\
\hline
$200$  & 0.205 & 0.254 & 1.006 & 0.911 & 0.954 & 3.112 \\
$500$  & 0.139 & 0.172 & 0.627 & 0.870 & 0.936 & 3.023 \\
$1000$ & 0.098 & 0.119 & 0.436 & 0.879 & 0.941 & 2.772 \\
\hline
\end{tabular}
\label{tab:KM_dr_ATE_I}
\end{table}

\begin{table}[H]
\centering
\small
\caption{Simulation results for the propensity-score kernel-matching ATE estimator under Setting II}
\begin{tabular}{ccccccc}
\hline
Sample size & MAE & RMSE & AW & 90\% Coverage & 95\% Coverage & NSD \\
\hline
$200$  & 0.355 & 0.443 & 1.787 & 0.901 & 0.964 & 6.259 \\
$500$  & 0.218 & 0.275 & 1.119 & 0.898 & 0.964 & 6.148 \\
$1000$ & 0.149 & 0.190 & 0.791 & 0.927 & 0.962 & 6.004 \\
\hline
\end{tabular}
\label{tab:KM_dr_ATE_II}
\end{table}

\begin{table}[H]
\centering
\small
\caption{Simulation results for the propensity-score kernel-matching ATT estimator under Setting I}
\begin{tabular}{ccccccc}
\hline
Sample size & MAE & RMSE & AW & 90\% Coverage & 95\% Coverage & NSD \\
\hline
$200$  & 0.164 & 0.206 & 0.835 & 0.919 & 0.961 & 2.919 \\
$500$  & 0.105 & 0.130 & 0.522 & 0.903 & 0.953 & 2.927 \\
$1000$ & 0.077 & 0.098 & 0.371 & 0.886 & 0.935 & 3.126 \\
\hline
\end{tabular}
\label{tab:KM_dr_ATT_I}
\end{table}

\begin{table}[H]
\centering
\small
\caption{Simulation results for the propensity-score kernel-matching ATT estimator under Setting II}
\begin{tabular}{ccccccc}
\hline
Sample size & MAE & RMSE & AW & 90\% Coverage & 95\% Coverage & NSD \\
\hline
$200$  & 0.463 & 0.581 & 2.318 & 0.893 & 0.945 & 8.230 \\
$500$  & 0.287 & 0.360 & 1.457 & 0.899 & 0.952 & 8.068 \\
$1000$ & 0.213 & 0.269 & 1.049 & 0.894 & 0.948 & 8.516 \\
\hline
\end{tabular}
\label{tab:KM_dr_ATT_II}
\end{table}

Tables \ref{tab:KM_dr_ATE_I}--\ref{tab:KM_dr_ATT_II} report the results
for the propensity-score kernel-matching estimators. For both
estimators, the MAE and RMSE decrease as the sample size increases, and
the empirical coverage probabilities are generally close to their
nominal coverage rates. The NSDs are also close to the square roots of the
corresponding efficiency bounds, particularly under Setting II.
Although some finite-sample deviations remain under Setting I, the
overall results are consistent with the efficiency result in
Proposition \ref{prop:ps-efficiency}.

\begin{table}[H]
\centering
\small
\caption{Simulation results for the 
doubly robust kernel-matching estimator of the ATE under Setting I}
\begin{tabular}{ccccccc}
\hline
Sample size & MAE & RMSE & AW & 90\% Coverage & 95\% Coverage & NSD \\
\hline
$200$  & 0.187 & 0.233 & 0.920 & 0.901 & 0.947 & 2.900 \\
$500$  & 0.121 & 0.148 & 0.577 & 0.893 & 0.959 & 2.718 \\
$1000$ & 0.081 & 0.100 & 0.409 & 0.922 & 0.968 & 2.485 \\
\hline
\end{tabular}
\label{tab:q1}
\end{table}

\begin{table}[H]
\centering
\small
\caption{Simulation results for the doubly robust kernel-matching estimator of the ATE under Setting II}
\begin{tabular}{ccccccc}
\hline
Sample size & MAE & RMSE & AW & 90\% Coverage & 95\% Coverage & NSD \\
\hline
$200$  & 0.345 & 0.434 & 1.690 & 0.890 & 0.943 & 6.434 \\
$500$  & 0.226 & 0.287 & 1.068 & 0.885 & 0.943 & 6.184 \\
$1000$ & 0.151 & 0.189 & 0.753 & 0.904 & 0.952 & 6.160 \\
\hline
\end{tabular}
\label{tab:q2}
\end{table}

\begin{table}[H]
\centering
\small
\caption{Simulation results for the doubly robust kernel-matching estimator of the ATT under Setting I}
\begin{tabular}{ccccccc}
\hline
Sample size & MAE & RMSE & AW & 90\% Coverage & 95\% Coverage & NSD \\
\hline
$200$  & 0.206 & 0.256 & 0.929 & 0.862 & 0.927 & 2.874 \\
$500$  & 0.130 & 0.159 & 0.593 & 0.883 & 0.945 & 2.811 \\
$1000$ & 0.076 & 0.095 & 0.426 & 0.944 & 0.971 & 2.748 \\
\hline
\end{tabular}
\label{tab:q3}
\end{table}

\begin{table}[H]
\centering
\small
\caption{Simulation results for the doubly robust kernel-matching estimator of the ATT under Setting II}
\begin{tabular}{ccccccc}
\hline
Sample size & MAE & RMSE & AW & 90\% Coverage & 95\% Coverage & NSD \\
\hline
$200$  & 0.530 & 0.697 & 2.094 & 0.882 & 0.939 & 9.851 \\
$500$  & 0.340 & 0.446 & 1.341 & 0.885 & 0.943 & 9.971 \\
$1000$ & 0.213 & 0.268 & 0.860 & 0.897 & 0.951 & 8.475 \\
\hline
\end{tabular}
\label{tab:q4}
\end{table}

Tables \ref{tab:q1}--\ref{tab:q4} present the corresponding results
for the doubly robust kernel-matching estimators. The MAE and RMSE
decrease steadily with the sample size, while the empirical coverage
probabilities are generally close to their nominal coverage rates. Moreover,
at \(N=1000\), the NSDs of the ATE estimators are \(2.485\) and
\(6.160\), compared with efficiency-bound values of \(2.473\) and
\(6.139\). The corresponding NSDs of the ATT estimators are \(2.748\)
and \(8.475\), compared with efficiency-bound values of \(2.816\) and
\(8.449\). Thus, the normalized variability of the estimators approaches
the corresponding efficiency bounds as the sample size increases.
Overall, the simulation results provide finite-sample evidence
consistent with the asymptotic efficiency conclusions in Theorems
\ref{th:ATEBOUND} and \ref{th:ATTBOUND}.

\section*{D. Simulation Results for Selection of Bandwidth and Kernel Functions}

\renewcommand{\thesection}{\arabic{section}}
\setcounter{equation}{0}
\renewcommand{\theequation}{D.\arabic{equation}}
\setcounter{table}{0}
\renewcommand{\thetable}{D.\arabic{table}}

\begin{figure}[H]
\centering
\renewcommand{\thefigure}{D.\arabic{figure}} 
\includegraphics[width=0.7\linewidth]{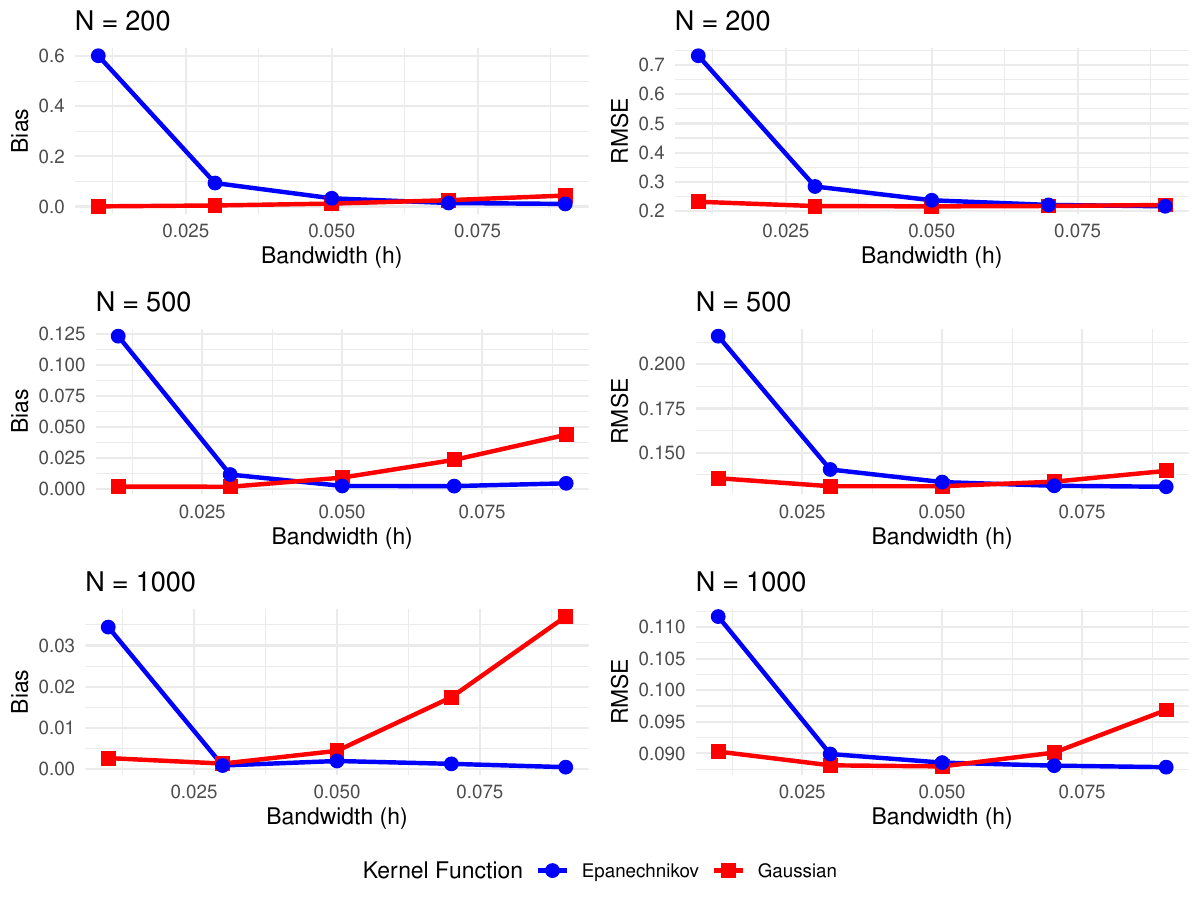}
\caption{Biases and RMSEs of ATT for Gaussian and Epanechnikov kernel matching across bandwidths ($h$) in Setting I ($N = 200, 500, 1000$).}
\label{fig:ATT1}
\end{figure}

\begin{figure}[H]
\centering
\renewcommand{\thefigure}{D.\arabic{figure}} 
\includegraphics[width=0.7\linewidth]{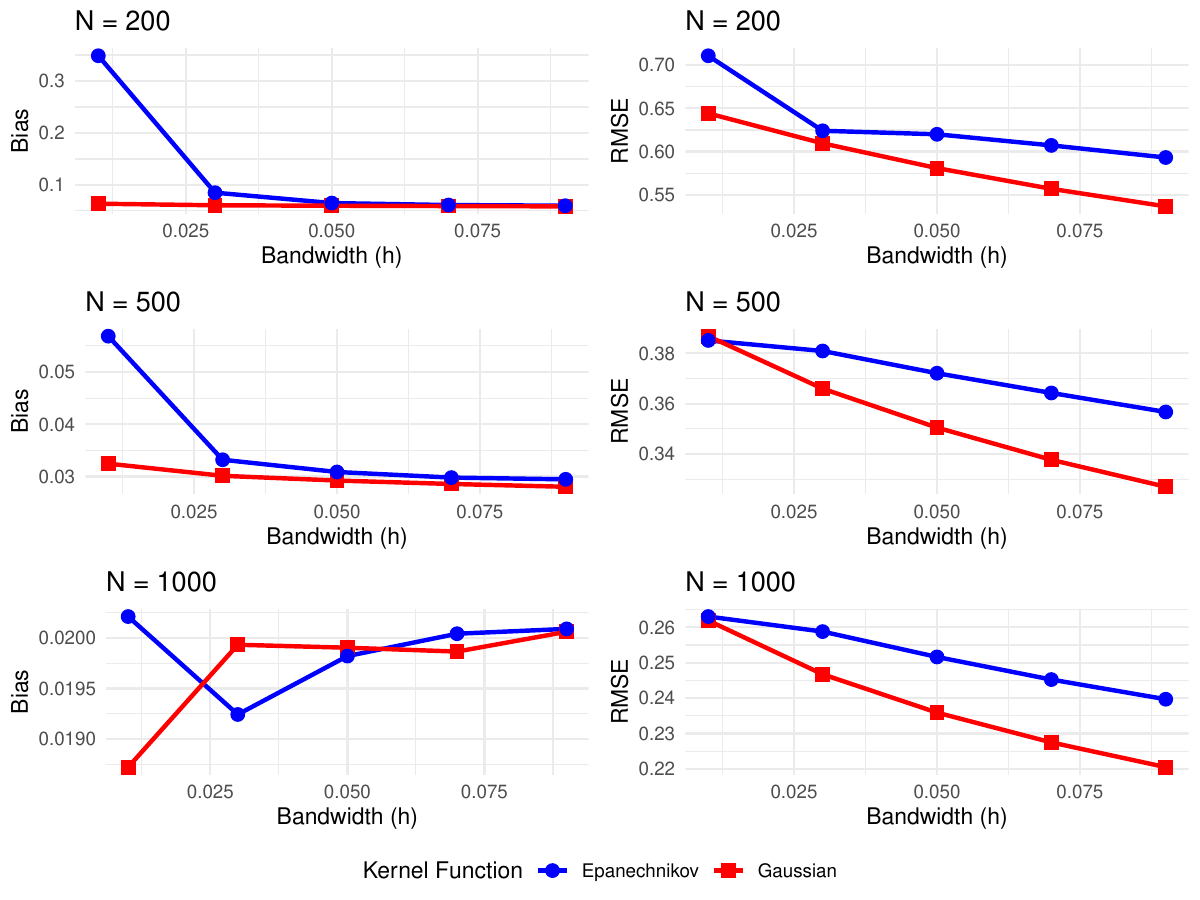}
\caption{Biases and RMSEs of ATT for Gaussian and Epanechnikov kernel matching across bandwidths ($h$) in Setting II ($N = 200, 500, 1000$).}
\label{fig:ATT2}
\end{figure}

\begin{figure}[H]
\centering
\renewcommand{\thefigure}{D.\arabic{figure}} 
\includegraphics[width=0.7\linewidth]{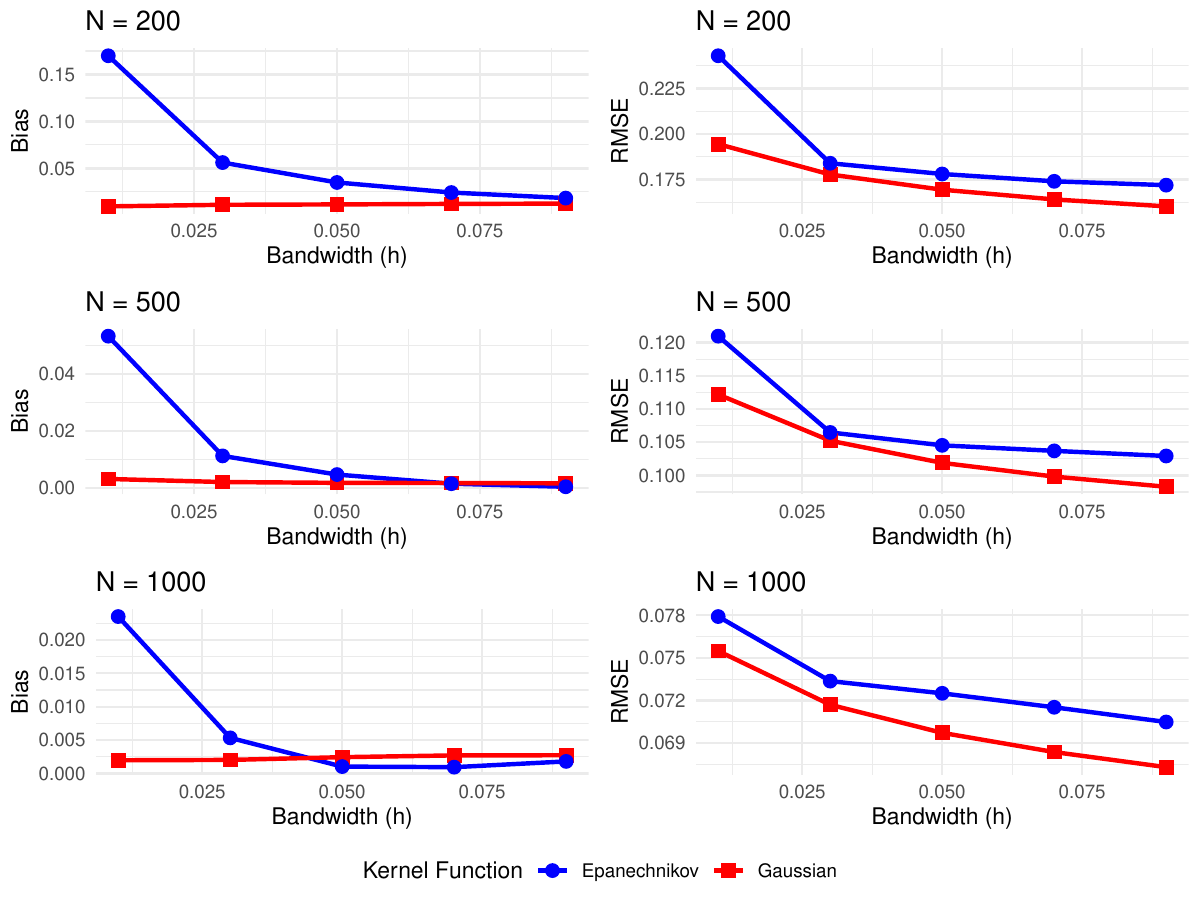}
\caption{Biases and RMSEs of ATT for Gaussian and Epanechnikov kernel matching across bandwidths ($h$) in Setting III ($N = 200, 500, 1000$).}
\label{fig:ATT3}
\end{figure}

\begin{figure}[H]
\centering
\renewcommand{\thefigure}{D.\arabic{figure}} 
\includegraphics[width=0.7\linewidth]{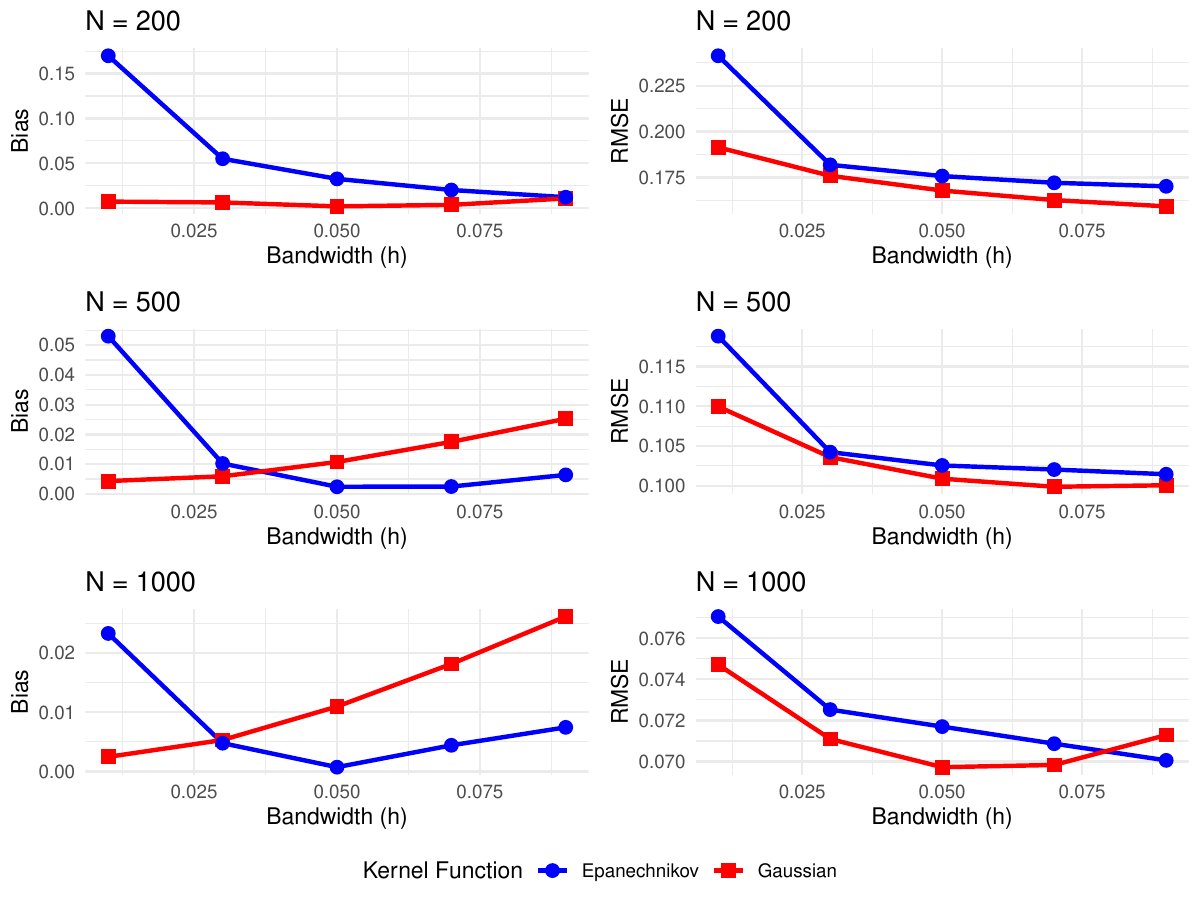}
\caption{Biases and RMSEs of ATT for Gaussian and Epanechnikov kernel matching across bandwidths ($h$) in Setting IV ($N = 200, 500, 1000$).}
\label{fig:ATT4}
\end{figure}

\begin{figure}[H]
\centering
\renewcommand{\thefigure}{D.\arabic{figure}} 
\includegraphics[width=0.7\linewidth]{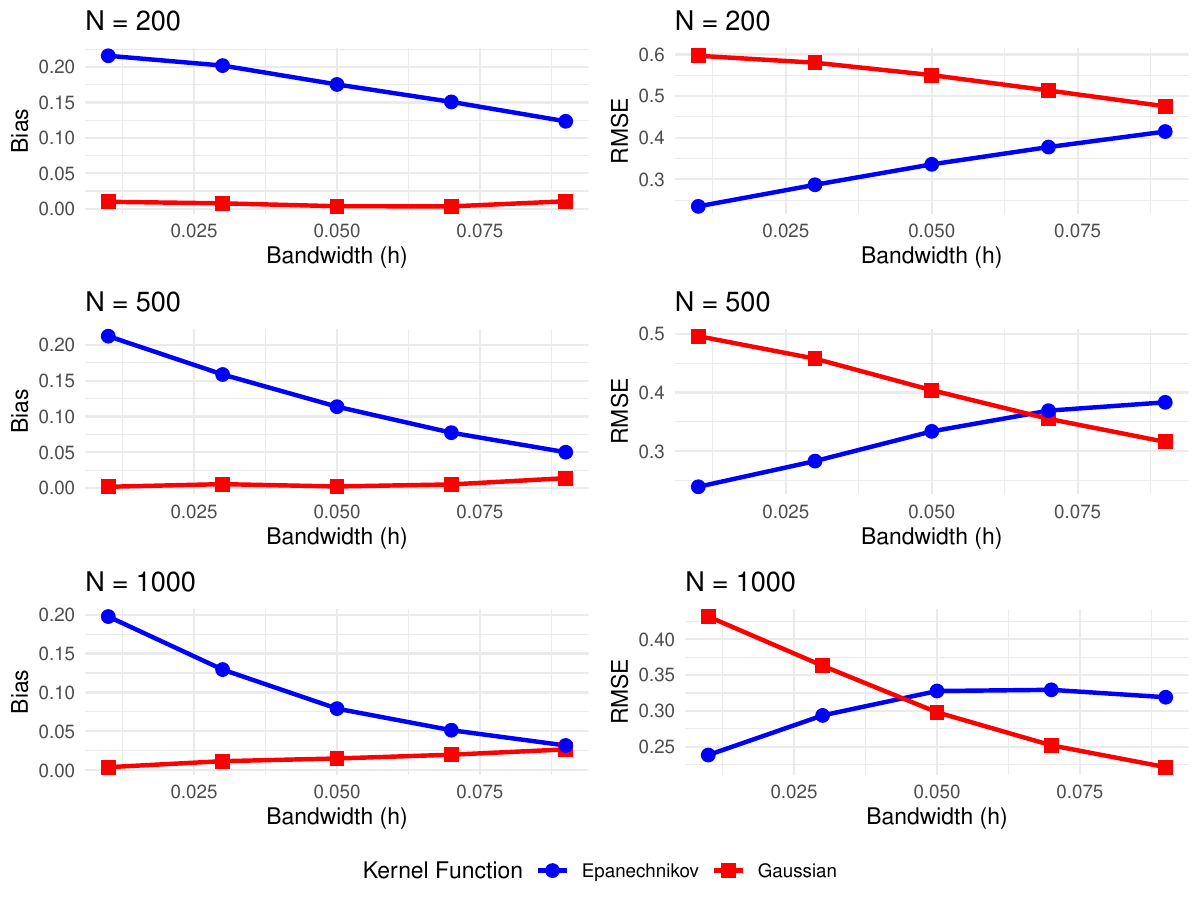}
\caption{Biases and RMSEs of ATT for Gaussian and Epanechnikov kernel matching across bandwidths ($h$) in Setting V ($N = 200, 500, 1000$).}
\label{fig:ATT5}
\vfill
\end{figure}
\clearpage

\section*{E. Simulation Results for Section \ref{sec:4.3}}

\renewcommand{\thesection}{\arabic{section}}
\setcounter{equation}{0}
\renewcommand{\theequation}{E.\arabic{equation}}
\setcounter{table}{0}
\renewcommand{\thetable}{E.\arabic{table}}

\begin{table}[H]
\centering
\caption{Simulation results for ATE in Setting I}
\label{tab:S1ATE}
\begin{tabular}{ccccccc}
\hline
Sample Size & Method        & \multicolumn{1}{r}{Bias} & SD     & RMSE   & AW    & CP    \\ \hline
            & Covariates    & -0.0042                  & 0.1963 & 0.1963 & \textcolor{red}{0.766} & \textcolor{red}{0.943} \\
            & True PS       & -0.0005                  & 0.2839 & 0.2839 & \textcolor{red}{1.080} & \textcolor{red}{0.943} \\
$N=200$     & Estimated PS  & 0.0024                   & 0.2315 & 0.2315 & \textcolor{red}{0.863} & \textcolor{red}{0.932} \\
            & Proposed      & -0.0040                  & 0.1986 & 0.1986 & \textcolor{red}{0.669} & \textcolor{red}{0.931} \\
            & IPW           & -0.0087                  & 0.1816 & 0.1818 & \textcolor{red}{0.943} & \textcolor{red}{0.913} \\
            & Doubly Robust & -0.0084                  & 0.1832 & 0.1834 & \textcolor{red}{0.888} & \textcolor{red}{0.958} \\ \hline
            & Covariates    & 0.0061                   & 0.1230 & 0.1232 & \textcolor{red}{0.484} & \textcolor{red}{0.939} \\
            & True PS       & -0.0023                  & 0.1749 & 0.1749 & \textcolor{red}{0.685} & \textcolor{red}{0.957} \\
$N=500$     & Estimated PS  & -0.0005                  & 0.1434 & 0.1434 & \textcolor{red}{0.555} & \textcolor{red}{0.958} \\
            & Proposed      & 0.0044                   & 0.1151 & 0.1152 & \textcolor{red}{0.438} & \textcolor{red}{0.945} \\
            & IPW           & 0.0011                   & 0.1143 & 0.1143 & \textcolor{red}{0.591} & \textcolor{red}{0.933} \\
            & Doubly Robust & 0.0010                   & 0.1129 & 0.1129 & \textcolor{red}{0.514} & \textcolor{red}{0.961} \\ \hline
            & Covariates    & 0.0037                   & 0.0835 & 0.0836 & \textcolor{red}{0.341} & \textcolor{red}{0.952} \\
            & True PS       & 0.0034                   & 0.1194 & 0.1194 & \textcolor{red}{0.485} & \textcolor{red}{0.948} \\
$N=1000$    & Estimated PS  & 0.0010                   & 0.0965 & 0.0965 & \textcolor{red}{0.396} & \textcolor{red}{0.949} \\
            & Proposed      & 0.0038                   & 0.0769 & 0.0770 & \textcolor{red}{0.313} & \textcolor{red}{0.950} \\
            & IPW           & 0.0003                   & 0.0768 & 0.0768 & \textcolor{red}{0.418} & \textcolor{red}{0.963} \\
            & Doubly Robust & 0.0012                   & 0.0769 & 0.0769 & \textcolor{red}{0.354} & \textcolor{red}{0.949} \\ \hline
\end{tabular}
\end{table}

\begin{table}[H]
\centering
\caption{Simulation results for ATE in Setting II}
\label{tab:S2ATE}
\begin{tabular}{ccccccc}
\hline
Sample size & Method        & Bias    & SD     & RMSE   & {\color{red} AW}    & {\color{red} CP}    \\ \hline
            & Covariates    & -0.0103 & 0.4438 & 0.4439 & {\color{red} 1.738} & {\color{red} 0.950} \\
            & True PS       & -0.0102 & 0.5329 & 0.5329 & {\color{red} 2.073} & {\color{red} 0.947} \\
$N=200$     & Estimated PS  & -0.0118 & 0.5189 & 0.5190 & {\color{red} 2.073} & {\color{red} 0.948} \\
            & Proposed      & -0.0068 & 0.4711 & 0.4712 & {\color{red} 1.752} & {\color{red} 0.948} \\
            & IPW           & -0.0073 & 0.4579 & 0.4580 & {\color{red} 1.715} & {\color{red} 0.938} \\
            & Doubly Robust & -0.0100 & 0.4609 & 0.4610 & {\color{red} 1.850} & {\color{red} 0.942} \\ \hline
            & Covariates    & -0.0028 & 0.2854 & 0.2854 & {\color{red} 1.095} & {\color{red} 0.940} \\
            & True PS       & -0.0018 & 0.3340 & 0.3340 & {\color{red} 1.314} & {\color{red} 0.948} \\
$N=500$     & Estimated PS  & -0.0166 & 0.3320 & 0.3324 & {\color{red} 1.313} & {\color{red} 0.949} \\
            & Proposed      & -0.0040 & 0.2958 & 0.2959 & {\color{red} 1.111} & {\color{red} 0.950} \\
            & IPW           & -0.0028 & 0.2950 & 0.2950 & {\color{red} 1.082} & {\color{red} 0.932} \\
            & Doubly Robust & -0.0018 & 0.2948 & 0.2948 & {\color{red} 1.147} & {\color{red} 0.939} \\ \hline
            & Covariates    & 0.0032  & 0.1891 & 0.1891 & {\color{red} 0.776} & {\color{red} 0.947} \\
            & True PS       & 0.0040  & 0.2245 & 0.2245 & {\color{red} 0.930} & {\color{red} 0.947} \\
$N=1000$    & Estimated PS  & 0.0104  & 0.2232 & 0.2235 & {\color{red} 0.930} & {\color{red} 0.948} \\
            & Proposed      & 0.0053  & 0.1954 & 0.1954 & {\color{red} 0.790} & {\color{red} 0.950} \\
            & IPW           & 0.0048  & 0.1930 & 0.1931 & {\color{red} 0.764} & {\color{red} 0.947} \\
            & Doubly Robust & 0.0044  & 0.1943 & 0.1944 & {\color{red} 0.806} & {\color{red} 0.953} \\ \hline
\end{tabular}
\end{table}

\begin{table}[H]
\centering
\caption{Simulation results for ATE in Setting III}
\label{tab:S3ATE}
\begin{tabular}{ccccccc}
\hline
Sample size & Method        & Bias    & SD     & RMSE   & {\color{red} AW}    & {\color{red} CP}    \\ \hline
            & Covariates    & -0.0092 & 0.1722 & 0.1724 & {\color{red} 0.888} & {\color{red} 0.940} \\
            & True PS       & -0.0103 & 0.1845 & 0.1848 & {\color{red} 1.273} & {\color{red} 0.939} \\
$N=200$     & Estimated PS  & -0.0089 & 0.1871 & 0.1873 & {\color{red} 1.173} & {\color{red} 0.938} \\
            & Proposed      & -0.0106 & 0.1726 & 0.1729 & {\color{red} 0.861} & {\color{red} 0.946} \\
            & IPW           & -0.0115 & 0.1601 & 0.1605 & {\color{red} 0.575} & {\color{red} 0.962} \\
            & Doubly Robust & -0.0105 & 0.1631 & 0.1635 & {\color{red} 0.731} & {\color{red} 0.955} \\ \hline
            & Covariates    & 0.0028  & 0.1088 & 0.1089 & {\color{red} 0.621} & {\color{red} 0.944} \\
            & True PS       & 0.0017  & 0.1166 & 0.1166 & {\color{red} 0.946} & {\color{red} 0.931} \\
$N=500$     & Estimated PS  & 0.0019  & 0.1147 & 0.1147 & {\color{red} 0.913} & {\color{red} 0.937} \\
            & Proposed      & 0.0011  & 0.1026 & 0.1026 & {\color{red} 0.656} & {\color{red} 0.953} \\
            & IPW           & 0.0014  & 0.0983 & 0.0983 & {\color{red} 0.760} & {\color{red} 0.953} \\
            & Doubly Robust & 0.0011  & 0.0987 & 0.0987 & {\color{red} 0.730} & {\color{red} 0.956} \\ \hline
            & Covariates    & 0.0032  & 0.0730 & 0.0731 & {\color{red} 0.473} & {\color{red} 0.956} \\
            & True PS       & 0.0004  & 0.0781 & 0.0781 & {\color{red} 0.746} & {\color{red} 0.949} \\
$N=1000$    & Estimated PS  & 0.0011  & 0.0770 & 0.0770 & {\color{red} 0.735} & {\color{red} 0.952} \\
            & Proposed      & 0.0016  & 0.0680 & 0.0680 & {\color{red} 0.537} & {\color{red} 0.950} \\
            & IPW           & 0.0010  & 0.0662 & 0.0662 & {\color{red} 0.554} & {\color{red} 0.957} \\
            & Doubly Robust & 0.0010  & 0.0669 & 0.0669 & {\color{red} 0.595} & {\color{red} 0.942} \\ \hline
\end{tabular}
\end{table}

\begin{table}[H]
\centering
\caption{Simulation results for ATE in Setting IV}
\label{tab:S4ATE}
\begin{tabular}{ccccccc}
\hline
Sample size & Method        & Bias    & SD     & RMSE   & {\color{red} AW}    & {\color{red} CP}    \\ \hline
            & Covariates    & 0.0328  & 0.1691 & 0.1722 & {\color{red} 0.659} & {\color{red} 0.945} \\
            & True PS       & -0.0082 & 0.1830 & 0.1832 & {\color{red} 0.709} & {\color{red} 0.958} \\
$N=200$     & Estimated PS  & -0.0066 & 0.1840 & 0.1842 & {\color{red} 0.696} & {\color{red} 0.936} \\
            & Proposed      & -0.0084 & 0.1708 & 0.1710 & {\color{red} 0.576} & {\color{red} 0.951} \\
            & IPW           & -0.0101 & 0.1589 & 0.1593 & {\color{red} 0.572} & {\color{red} 0.885} \\
            & Doubly Robust & -0.0098 & 0.1655 & 0.1658 & {\color{red} 0.775} & {\color{red} 0.958} \\ \hline
            & Covariates    & 0.0326  & 0.1067 & 0.1116 & {\color{red} 0.418} & {\color{red} 0.929} \\
            & True PS       & 0.0024  & 0.1141 & 0.1141 & {\color{red} 0.451} & {\color{red} 0.966} \\
$N=500$     & Estimated PS  & 0.0029  & 0.1126 & 0.1127 & {\color{red} 0.448} & {\color{red} 0.953} \\
            & Proposed      & 0.0024  & 0.1005 & 0.1006 & {\color{red} 0.375} & {\color{red} 0.950} \\
            & IPW           & 0.0018  & 0.0967 & 0.0967 & {\color{red} 0.359} & {\color{red} 0.926} \\
            & Doubly Robust & 0.0014  & 0.0994 & 0.0994 & {\color{red} 0.449} & {\color{red} 0.960} \\ \hline
            & Covariates    & 0.0262  & 0.0718 & 0.0765 & {\color{red} 0.297} & {\color{red} 0.931} \\
            & True PS       & 0.0008  & 0.0767 & 0.0767 & {\color{red} 0.319} & {\color{red} 0.942} \\
$N=1000$    & Estimated PS  & 0.0014  & 0.0757 & 0.0757 & {\color{red} 0.319} & {\color{red} 0.940} \\
            & Proposed      & 0.0022  & 0.0671 & 0.0671 & {\color{red} 0.269} & {\color{red} 0.951} \\
            & IPW           & 0.0012  & 0.0652 & 0.0652 & {\color{red} 0.278} & {\color{red} 0.934} \\
            & Doubly Robust & 0.0013  & 0.0678 & 0.0678 & {\color{red} 0.304} & {\color{red} 0.945} \\ \hline
\end{tabular}
\end{table}

\begin{table}[H]
\centering
\caption{Simulation results for ATE in Setting V}
\label{tab:S5ATE}
\begin{tabular}{ccccccc}
\hline
Sample size & Method        & Bias    & SD     & RMSE   & {\color{red} AW}    & {\color{red} CP}    \\ \hline
            & Covariates    & -0.0009 & 0.3426 & 0.3426 & {\color{red} 1.384} & {\color{red} 0.930} \\
            & True PS       & -0.0072 & 0.6568 & 0.6568 & {\color{red} 2.143} & {\color{red} 0.965} \\
$N=200$     & Estimated PS  & -0.0178 & 0.6552 & 0.6555 & {\color{red} 1.832} & {\color{red} 0.965} \\
            & Proposed      & -0.0059 & 0.5001 & 0.5001 & {\color{red} 1.204} & {\color{red} 0.959} \\
            & IPW           & 0.0218  & 0.4854 & 0.4859 & {\color{red} 2.045} & {\color{red} 0.899} \\
            & Doubly Robust & 0.2714  & 2.4132 & 2.4284 & {\color{red} 3.837} & {\color{red} 0.756} \\ \hline
            & Covariates    & -0.0113 & 0.2713 & 0.2715 & {\color{red} 1.077} & {\color{red} 0.939} \\
            & True PS       & -0.0410 & 0.5193 & 0.5209 & {\color{red} 1.788} & {\color{red} 0.971} \\
$N=500$     & Estimated PS  & -0.0469 & 0.5095 & 0.5116 & {\color{red} 1.678} & {\color{red} 0.973} \\
            & Proposed      & 0.0007  & 0.4088 & 0.4088 & {\color{red} 1.013} & {\color{red} 0.956} \\
            & IPW           & 0.0101  & 0.3903 & 0.3904 & {\color{red} 3.147} & {\color{red} 0.733} \\
            & Doubly Robust & 0.2610  & 3.7630 & 3.7720 & {\color{red} 2.268} & {\color{red} 0.676} \\ \hline
            & Covariates    & -0.0196 & 0.2296 & 0.2305 & {\color{red} 0.889} & {\color{red} 0.979} \\
            & True PS       & 0.0098  & 0.4852 & 0.4853 & {\color{red} 1.514} & {\color{red} 0.971} \\
$N=1000$    & Estimated PS  & 0.0069  & 0.4826 & 0.4827 & {\color{red} 1.480} & {\color{red} 0.969} \\
            & Proposed      & 0.0042  & 0.3545 & 0.3545 & {\color{red} 0.892} & {\color{red} 0.953} \\
            & IPW           & 0.0119  & 0.3653 & 0.3655 & {\color{red} 1.305} & {\color{red} 0.729} \\
            & Doubly Robust & 0.0403  & 3.1647 & 3.1650 & {\color{red} 2.213} & {\color{red} 0.697} \\ \hline
\end{tabular}
\end{table}

\begin{table}[H]
\centering
\caption{Simulation results for ATT in Setting I}
\label{tab:S1ATT}
\begin{tabular}{ccccccc}
\hline
Sample size & Method        & Bias    & SD     & RMSE   & {\color{red} AW}    & {\color{red} CP}    \\ \hline
            & Covariates    & -0.0136 & 0.2352 & 0.2355 & {\color{red} 0.915} & {\color{red} 0.929} \\
            & True PS       & 0.0052  & 0.3396 & 0.3394 & {\color{red} 1.258} & {\color{red} 0.935} \\
$N=200$     & Estimated PS  & 0.0105  & 0.2836 & 0.2836 & {\color{red} 1.079} & {\color{red} 0.925} \\
            & Proposed      & -0.0247 & 0.2161 & 0.2174 & {\color{red} 0.783} & {\color{red} 0.939} \\
            & IPW           & -0.0029 & 0.2100 & 0.2099 & {\color{red} 0.811} & {\color{red} 0.943} \\
            & Doubly Robust & -0.0043 & 0.2045 & 0.2044 & {\color{red} 0.932} & {\color{red} 0.942} \\ \hline
            & Covariates    & -0.0048 & 0.1465 & 0.1465 & {\color{red} 0.576} & {\color{red} 0.940} \\
            & True PS       & -0.0099 & 0.2071 & 0.2072 & {\color{red} 0.805} & {\color{red} 0.960} \\
$N=500$     & Estimated PS  & -0.0008 & 0.1796 & 0.1795 & {\color{red} 0.696} & {\color{red} 0.938} \\
            & Proposed      & -0.0120 & 0.1324 & 0.1329 & {\color{red} 0.509} & {\color{red} 0.946} \\
            & IPW           & -0.0038 & 0.1303 & 0.1303 & {\color{red} 0.506} & {\color{red} 0.946} \\
            & Doubly Robust & -0.0050 & 0.1269 & 0.1269 & {\color{red} 0.587} & {\color{red} 0.973} \\ \hline
            & Covariates    & 0.0005  & 0.1040 & 0.1040 & {\color{red} 0.408} & {\color{red} 0.959} \\
            & True PS       & 0.0027  & 0.1485 & 0.1485 & {\color{red} 0.569} & {\color{red} 0.951} \\
$N=1000$    & Estimated PS  & 0.0026  & 0.1231 & 0.1230 & {\color{red} 0.497} & {\color{red} 0.953} \\
            & Proposed      & 0.0018  & 0.0923 & 0.0923 & {\color{red} 0.363} & {\color{red} 0.956} \\
            & IPW           & 0.0021  & 0.0911 & 0.0911 & {\color{red} 0.358} & {\color{red} 0.957} \\
            & Doubly Robust & 0.0004  & 0.0881 & 0.0880 & {\color{red} 0.415} & {\color{red} 0.978} \\ \hline
\end{tabular}
\end{table}

\begin{table}[H]
\centering
\caption{Simulation results for ATT in Setting II}
\label{tab:S2ATT}
\begin{tabular}{ccccccc}
\hline
Sample size & Method        & Bias    & SD     & RMSE   & {\color{red} AW}    & {\color{red} CP}    \\ \hline
            & Covariates    & -0.0670 & 0.6170 & 0.6203 & {\color{red} 2.379} & {\color{red} 0.937} \\
            & True PS       & -0.0691 & 0.6115 & 0.6151 & {\color{red} 2.391} & {\color{red} 0.945} \\
$N=200$     & Estimated PS  & -0.0802 & 0.7186 & 0.7227 & {\color{red} 2.391} & {\color{red} 0.945} \\
            & Proposed      & -0.0593 & 0.5541 & 0.5570 & {\color{red} 1.802} & {\color{red} 0.950} \\
            & IPW           & -0.0669 & 0.6285 & 0.6317 & {\color{red} 1.842} & {\color{red} 0.947} \\
            & Doubly Robust & -0.0649 & 0.6133 & 0.6164 & {\color{red} 2.471} & {\color{red} 0.968} \\ \hline
            & Covariates    & -0.0240 & 0.3789 & 0.3795 & {\color{red} 1.507} & {\color{red} 0.946} \\
            & True PS       & -0.0524 & 0.3789 & 0.3823 & {\color{red} 1.513} & {\color{red} 0.953} \\
$N=500$     & Estimated PS  & -0.0449 & 0.4442 & 0.4462 & {\color{red} 1.512} & {\color{red} 0.954} \\
            & Proposed      & -0.0339 & 0.3427 & 0.3442 & {\color{red} 1.147} & {\color{red} 0.951} \\
            & IPW           & -0.0315 & 0.3813 & 0.3825 & {\color{red} 1.161} & {\color{red} 0.943} \\
            & Doubly Robust & -0.0301 & 0.3761 & 0.3771 & {\color{red} 1.544} & {\color{red} 0.966} \\ \hline
            & Covariates    & -0.0268 & 0.2725 & 0.2736 & {\color{red} 1.067} & {\color{red} 0.952} \\
            & True PS       & -0.0280 & 0.2672 & 0.2685 & {\color{red} 1.072} & {\color{red} 0.943} \\
$N=1000$    & Estimated PS  & -0.0185 & 0.3126 & 0.3130 & {\color{red} 1.072} & {\color{red} 0.943} \\
            & Proposed      & -0.0260 & 0.2544 & 0.2556 & {\color{red} 0.82}  & {\color{red} 0.944} \\
            & IPW           & -0.0272 & 0.2748 & 0.2760 & {\color{red} 0.822} & {\color{red} 0.946} \\
            & Doubly Robust & -0.0286 & 0.2690 & 0.2704 & {\color{red} 1.089} & {\color{red} 0.969} \\ \hline
\end{tabular}
\end{table}

\begin{table}[H]
\centering
\caption{Simulation results for ATT in Setting III}
\label{tab:S3ATt}
\begin{tabular}{ccccccc}
\hline
Sample size & Method        & Bias    & SD     & RMSE   & \textcolor{red}{AW}    & \textcolor{red}{CP}    \\ \hline
            & Covariates    & -0.0060 & 0.1951 & 0.1951 & \textcolor{red}{1.082} & \textcolor{red}{0.937} \\
            & True PS       & -0.0041 & 0.2183 & 0.2182 & \textcolor{red}{1.566} & \textcolor{red}{0.864} \\
$N=200$     & Estimated PS  & -0.0045 & 0.2194 & 0.2193 & \textcolor{red}{1.451} & \textcolor{red}{0.807} \\
            & Proposed      & -0.0122 & 0.1636 & 0.1640 & \textcolor{red}{1.062} & \textcolor{red}{0.911} \\
            & IPW           & -0.0135 & 0.1714 & 0.1718 & \textcolor{red}{1.424} & \textcolor{red}{0.871} \\
            & Doubly Robust & -0.0142 & 0.1718 & 0.1723 & \textcolor{red}{0.576} & \textcolor{red}{0.902} \\ \hline
            & Covariates    & 0.0039  & 0.1253 & 0.1253 & \textcolor{red}{0.79}  & \textcolor{red}{0.932} \\
            & True PS       & 0.0019  & 0.1401 & 0.1400 & \textcolor{red}{1.236} & \textcolor{red}{0.891} \\
$N=500$     & Estimated PS  & 0.0069  & 0.1372 & 0.1373 & \textcolor{red}{1.198} & \textcolor{red}{0.873} \\
            & Proposed      & 0.0038  & 0.1040 & 0.1040 & \textcolor{red}{0.847} & \textcolor{red}{0.943} \\
            & IPW           & 0.0041  & 0.1075 & 0.1076 & \textcolor{red}{1.025} & \textcolor{red}{0.893} \\
            & Doubly Robust & 0.0041  & 0.1082 & 0.1082 & \textcolor{red}{0.360} & \textcolor{red}{0.909} \\ \hline
            & Covariates    & 0.0040  & 0.0855 & 0.0854 & \textcolor{red}{0.617} & \textcolor{red}{0.939} \\
            & True PS       & -0.0030 & 0.0950 & 0.0950 & \textcolor{red}{0.967} & \textcolor{red}{0.891} \\
$N=1000$    & Estimated PS  & 0.0002  & 0.0935 & 0.0935 & \textcolor{red}{0.956} & \textcolor{red}{0.887} \\
            & Proposed      & -0.0004 & 0.0749 & 0.0749 & \textcolor{red}{0.697} & \textcolor{red}{0.942} \\
            & IPW           & -0.0009 & 0.0755 & 0.0755 & \textcolor{red}{0.833} & \textcolor{red}{0.903} \\
            & Doubly Robust & -0.0007 & 0.0756 & 0.0756 & \textcolor{red}{0.252} & \textcolor{red}{0.917} \\ \hline
\end{tabular}
\end{table}

\begin{table}[H]
\centering
\caption{Simulation results for ATT in Setting IV}
\label{tab:S4ATT}
\begin{tabular}{ccccccc}
\hline
Sample size & Method        & Bias    & SD     & RMSE   & {\color{red} AW}    & {\color{red} CP}    \\ \hline
            & Covariates    & 0.0360  & 0.1922 & 0.1954 & {\color{red} 0.740} & {\color{red} 0.940} \\
            & True PS       & -0.0019 & 0.2177 & 0.2176 & {\color{red} 0.825} & {\color{red} 0.941} \\
$N=200$     & Estimated PS  & -0.0021 & 0.2154 & 0.2153 & {\color{red} 0.834} & {\color{red} 0.946} \\
            & Proposed      & 0.0038  & 0.1627 & 0.1627 & {\color{red} 0.633} & {\color{red} 0.951} \\
            & IPW           & -0.0120 & 0.1709 & 0.1712 & {\color{red} 0.670} & {\color{red} 0.951} \\
            & Doubly Robust & -0.0140 & 0.1717 & 0.1721 & {\color{red} 0.573} & {\color{red} 0.947} \\ \hline
            & Covariates    & 0.0334  & 0.1236 & 0.1280 & {\color{red} 0.473} & {\color{red} 0.942} \\
            & True PS       & 0.0028  & 0.1359 & 0.1359 & {\color{red} 0.533} & {\color{red} 0.951} \\
$N=500$     & Estimated PS  & 0.0070  & 0.1337 & 0.1339 & {\color{red} 0.532} & {\color{red} 0.946} \\
            & Proposed      & 0.0121  & 0.1021 & 0.1028 & {\color{red} 0.404} & {\color{red} 0.957} \\
            & IPW           & 0.0040  & 0.1058 & 0.1058 & {\color{red} 0.422} & {\color{red} 0.956} \\
            & Doubly Robust & 0.0037  & 0.1064 & 0.1064 & {\color{red} 0.391} & {\color{red} 0.936} \\ \hline
            & Covariates    & 0.0236  & 0.0848 & 0.0880 & {\color{red} 0.337} & {\color{red} 0.942} \\
            & True PS       & -0.0023 & 0.0945 & 0.0945 & {\color{red} 0.376} & {\color{red} 0.953} \\
$N=1000$    & Estimated PS  & 0.0007  & 0.0932 & 0.0931 & {\color{red} 0.377} & {\color{red} 0.940} \\
            & Proposed      & 0.0033  & 0.0747 & 0.0747 & {\color{red} 0.292} & {\color{red} 0.951} \\
            & IPW           & -0.0007 & 0.0753 & 0.0752 & {\color{red} 0.298} & {\color{red} 0.933} \\
            & Doubly Robust & -0.0006 & 0.0750 & 0.0750 & {\color{red} 0.251} & {\color{red} 0.932} \\ \hline
\end{tabular}
\end{table}

\begin{table}[H]
\centering
\caption{Simulation results for ATT in Setting V}
\label{tab:S5ATT}
\begin{tabular}{ccccccc}
\hline
Sample size & Method        & Bias    & SD     & RMSE   & {\color{red} AW}    & {\color{red} CP}    \\ \hline
            & Covariates    & -0.0118 & 0.3964 & 0.3963 & {\color{red} 1.595} & {\color{red} 0.922} \\
            & True PS       & -0.0140 & 0.7851 & 0.7848 & {\color{red} 2.519} & {\color{red} 0.827} \\
$N=200$     & Estimated PS  & -0.0252 & 0.7820 & 0.7820 & {\color{red} 2.156} & {\color{red} 0.721} \\
            & Proposed      & 0.0032  & 0.5138 & 0.5135 & {\color{red} 1.371} & {\color{red} 0.925} \\
            & IPW           & 0.0021  & 0.5759 & 0.5756 & {\color{red} 2.286} & {\color{red} 0.562} \\
            & Doubly Robust & -0.1692 & 1.7912 & 1.7983 & {\color{red} 1.725} & {\color{red} 0.890} \\ \hline
            & Covariates    & -0.0006 & 0.3345 & 0.3343 & {\color{red} 1.271} & {\color{red} 0.936} \\
            & True PS       & -0.0272 & 0.6483 & 0.6486 & {\color{red} 2.140} & {\color{red} 0.830} \\
$N=500$     & Estimated PS  & -0.0255 & 0.6382 & 0.6383 & {\color{red} 2.011} & {\color{red} 0.784} \\
            & Proposed      & 0.0300  & 0.4181 & 0.4189 & {\color{red} 1.181} & {\color{red} 0.912} \\
            & IPW           & 0.0069  & 0.4691 & 0.4689 & {\color{red} 3.771} & {\color{red} 0.604} \\
            & Doubly Robust & -0.0203 & 3.1213 & 3.1198 & {\color{red} 1.889} & {\color{red} 0.921} \\ \hline
            & Covariates    & -0.0346 & 0.2742 & 0.2763 & {\color{red} 1.062} & {\color{red} 0.920} \\
            & True PS       & -0.0131 & 0.5775 & 0.5774 & {\color{red} 1.828} & {\color{red} 0.842} \\
$N=1000$    & Estimated PS  & -0.0120 & 0.5694 & 0.5692 & {\color{red} 1.786} & {\color{red} 0.799} \\
            & Proposed      & -0.0099 & 0.3616 & 0.3615 & {\color{red} 1.055} & {\color{red} 0.945} \\
            & IPW           & -0.0119 & 0.3862 & 0.3862 & {\color{red} 1.517} & {\color{red} 0.664} \\
            & Doubly Robust & -0.1516 & 0.6273 & 0.6450 & {\color{red} 2.170} & {\color{red} 0.939} \\ \hline
\end{tabular}
\end{table}

\end{document}